\documentclass[prd,aps,preprint,tightenlines,superscriptaddress,nofootinbib,showpacs,showkeys]{revtex4}
\usepackage{float}
\usepackage{mathrsfs}
\usepackage{amsfonts}
\usepackage{array}
\usepackage{epsfig}
\usepackage{amsmath}    
\usepackage{amssymb}   
\usepackage{graphicx}   
\usepackage{verbatim}   
\usepackage{color}      
\usepackage{subfigure}  
\usepackage{hyperref}   

\begin{document}

\title{
Intrinsic Charm Parton Distribution Functions \\
from CTEQ-TEA Global Analysis}

\author{Sayipjamal Dulat}
\email{sdulat@msu.edu}
\affiliation{
School of Physics Science and Technology, Xinjiang University,\\
 Urumqi, Xinjiang 830046 China }
\affiliation{
Department of Physics and Astronomy, Michigan State University,\\
 East Lansing, MI 48824 U.S.A. }
\author{Tie-Jiun Hou}
\affiliation{
Institute of Physics, Academia Sinica, Taipei, Taiwan 115 }
\author{Jun Gao}
\affiliation{
Department of Physics, Southern Methodist University,\\
 Dallas, TX 75275-0181, U.S.A. }
\author{Joey Huston}
\affiliation{
Department of Physics and Astronomy, Michigan State University,\\
 East Lansing, MI 48824 U.S.A. }
\author{Jon Pumplin}
\affiliation{
Department of Physics and Astronomy, Michigan State University,\\
 East Lansing, MI 48824 U.S.A. }
\author{Carl Schmidt}
\affiliation{
Department of Physics and Astronomy, Michigan State University,\\
 East Lansing, MI 48824 U.S.A. }
\author{Daniel Stump}
\affiliation{
Department of Physics and Astronomy, Michigan State University,\\
 East Lansing, MI 48824 U.S.A. }
\author{ C.--P. Yuan}
\email{yuan@pa.msu.edu}
\affiliation{
Department of Physics and Astronomy, Michigan State University,\\
 East Lansing, MI 48824 U.S.A. }

\begin{abstract}
We study the possibility of intrinsic (non-perturbative) charm in parton
 distribution functions (PDF) of the proton, within the context of 
 the CT10 next-to-next-to-leading order (NNLO) global analysis.
Three models for the intrinsic charm (IC) quark content
are compared:
(i) $\hat{c}(x) = 0$ (zero-IC model);
(ii) $\hat{c}(x)$ is parametrized by a valence-like parton distribution
(BHPS model);
(iii) $\hat{c}(x)$ is parametrized by a sea-like parton distribution
(SEA model).  In these models, the intrinsic charm content, $\hat{c}(x)$, is
included in the charm PDF at the matching scale $Q_c=m_c=1.3$ GeV.
The best fits to data are constructed and compared.
Correlations between the value of $m_c$ and the amount of IC are also
considered.


\end{abstract}



\maketitle

\section{Charmed quarks in the CTEQ-TEA global analysis
\label{sec:INTRODUCTION}}

The Global Analysis uses QCD theory to analyze a broad
range of experimental data.  In particular, the theoretical predictions
for short-distance scattering processes allow the measurement, within
some approximations, of universal parton distribution functions (PDFs)
for the proton.
These functions can then be used to predict hadronic cross sections in the
QCD and electroweak theories, and in beyond-the-standard-model theories.
With the new high-precision data becoming available from the LHC,
the goal of QCD global analysis is to be able to make predictions that
are accurate to be less than about one percent.

The most recent CT10NNLO PDFs
(referred to as the CT10 PDFs in this paper)
are based on next-to-next-to-leading order
(NNLO) approximations for perturbative QCD~\cite{ct10nn}.
That is, NNLO approximations are used for the running coupling
$\alpha_{\rm S}(Q)$,
for the DGLAP evolution equations,
and for those hard matrix elements for which the
NNLO approximation is available~\cite{NNLO:disme,NNLO:vbp}.(NLO is
 used only for inclusive jet data.)

Another important approximation in the CT10 analysis
concerns the treatment of charmed quark effects.
There are two issues:
the dependence on the assumed charmed quark mass,
and the possibility of a nonperturbative charm component
in the proton (intrinsic charm IC).
The first issue has been addressed in many
papers~\cite{acot,acotchi,acotnn,buza98,chuvakin00,thorne98,forte10,ball11}
and was considered recently in the context of the CT10 analysis
in Ref.~\cite{smucharm}.  In that work the general dependence on the
charm quark mass  was studied and a
preferred value of $m_c(m_c) = 1.15^{+0.18}_{-0.12}$ GeV
was obtained at 68\% C.L.,
where the error is a sum in quadrature of PDF and theoretical uncertainties.
Here, $m_c(m_c)$ denotes the running mass of the charm quark, defined in the
modified minimal-subtraction ($\overline {\rm MS}$) scheme and
evaluated at the scale of $m_c$.
This value, constrained primarily by a combination of inclusive and charm
production measurements in HERA deep-inelastic scattering,
translates  into $m_c^{\rm pole}=1.31^{+0.19}_{-0.13}$ GeV and
$1.54^{+0.18}_{-0.12}$ GeV if using the conversion
formula in Eq.~(17) of Ref.~\cite{rundec} at one and two loops.
Either converted value is compatible with the value of
$m_c^{\rm pole} = 1.3$ GeV, which  was assumed by CT10,
and which we shall use as our standard charm mass value in this paper.

The second issue, the possibility of intrinsic charm, is our primary concern here.
In the standard CT10 analysis, the charm and anti-charm quark PDFs were
turned on
at the scale $Q=Q_c=m_c=1.3$ GeV with an initial ${\cal O}(\alpha_s^2)$ distribution, consistent with
NNLO matching~\cite{NNLOmatching}.  Thus, at higher $Q$, most of the charm PDF is generated from DGLAP evolution.
However, one can also consider the possibility of including an additional contribution, $\hat{c}(x)$,
to the initial charm and anti-charm PDFs at the scale $Q_c$, beyond that required by matching.
In principle, this intrinsic charm content would be suppressed by powers of
$(\Lambda_{\rm QCD}/m_c)$, but since this ratio is not very small,
it may be important.

Thus, in this paper we ask these questions:
What are the nonperturbative $c$ and $\overline{c}$
components of the proton?
Is intrinsic charm significant?
Accurate predictions of the $c$ and $\overline{c}$ parton distributions
will be relevant to some important LHC measurements.
For example, production of $W^{\pm}$ and $Z^{0}$  involves
$c\overline{d}$, $c\overline{s}$, $d\overline{c}$, $s\overline{c}$
and $c\overline{c}$ contributions.
Another example is charmed particle production at the LHC,
which will depend quite directly on the $c$ and $\overline{c}$ partons;
some data for this process have already been published~\cite{lhccharm}.
In addition, because of the momentum sum rule,
an increase in the charm component of the proton must
be compensated by a decrease in other components.
So, in a model with IC, a compensating change in the gluon PDF
could change the theoretical predictions of other processes,
such as jet production at the Tevatron and Higgs boson production at the
LHC,
which are not directly related to charm.

The current paper updates previous work on the charm PDFs,
which was based on the CTEQ6.5 global analysis~\cite{plt}.
However, there are some important advances with respect
to the previous work.
Most importantly, the PDFs in this paper will be based
on the NNLO approximation of perturbative QCD;
the previous work was NLO.
Also, some more recent data is now available and is used here:
the \emph{combined} H1 and ZEUS data for both inclusive deep-inelastic
scattering~\cite{:2009wt}  and inclusive charm production~\cite{HERAX0c}  at HERA.
Given these improvements to theory and experiment, we expect that
this updated analysis will yield a better understanding of
the charm PDF.

The outline of the paper is as follows.
In Section 2, we present three models of the intrinsic charm quark
PDF, and obtain fits and constraints on these models
by performing a global analysis of the data, using a charm mass\footnote{%
Throughout this paper, unless otherwise specified, the variable
$m_c$ will indicate the charm mass in the
pole mass scheme.}
 of $m_c=1.3$ GeV.
In Section 3 we list the data used in the global analysis
and present comparisons between theory and data for the
data sets that are particularly sensitive to the charm quark PDF.
In Section 4 we use these PDFs
(with the corresponding PDFs for other partons, of course)
in calculations of  processes at the LHC
that would be particularly sensitive to the charm quark PDF.
In Section 5 we consider the correlation of the amount of intrinsic charm
with the value of the charm mass.  In particular, we demonstrate this correlation
by re-doing the fit for one of the models with the larger value of the charm mass
$m_c=1.67$ GeV.
Section 6 contains our conclusions. We also include an
appendix, which contains the technical definition
of a novel variable that we use to assess the comparison
of theory and data for the individual data sets.

\section{Fits with an Intrinsic Charm component
\label{sec:RESULTS}}

Following our earlier study of $c$ and $\overline{c}$
content of the proton~\cite{plt},
we will consider three models for the charm PDF $c(x,Q_c)$.
In all three models, the charm PDF becomes active at the
matching scale $Q_c=m_c=1.3$ GeV.
The first model is the standard CT10 PDFs;
they have an initial value of $c(x,Q_c)=\bar{c}(x,Q_c)$ that
 is fixed by NNLO matching\footnote{%
Note that, in the absence of intrinsic charm, the condition $c(x,Q_{c}) = \overline{c}(x,Q_{c}) = 0$
is only valid for $Q_c=m_c$ at NLO;  at NNLO,
the starting charm quark PDFs are
${\cal O}(\alpha_s^2)$ and nonzero, even for $Q_c=m_c$.}~\cite{NNLOmatching}
and is ${\cal O}(\alpha_s^2)$.
For $Q>Q_c$, additional charm content is generated by radiation,
as required by the evolution equations of the renormalization group.
For most values of $Q$ (larger than a few GeV)
this radiated charm content dominates the ${\cal O}(\alpha_s^2)$ initial contribution.

In the other two models, we assume some additional nonperturbative intrinsic charm (IC) content, $\hat{c}(x)$, which is
added to the ${\cal O}(\alpha_s^2)$ perturbative contribution at the scale $Q_c$.
The second model, which we call the BHPS model, has an IC content that is parametrized by
a ``valence-like'' nonperturbative function,
\begin{equation}\label{eq:modelB}
\hat{c}(x) =
A ~ x^{2}\left[ 6x(1+x){\rm ln}\,x + (1-x)(1+10x+x^{2}) \right] .
\end{equation}
This model, which is representative of predictions from the light-cone picture of nucleon structure,
is based on the original work of Brodsky, Hoyer, Peterson, and Sakai~\cite{bhps};
it was also considered in the NLO CTEQ6.5 study~\cite{plt}.
The third model, which we call the SEA model, has an IC content that is parametrized by a ``sea-like'' nonperturbative function,
\begin{equation}\label{eq:modelS}
\hat{c}(x)  = A ~
\left( \overline{d}(x,Q_{0})+\overline{u}(x,Q_{0}) \right) .
\end{equation}
For all three models we use the Fortran 95 package HOPPET~\cite{hoppet} to include the IC with the NNLO matching
and to evolve the PDFs at NNLO.  We also use the NNLO S-ACOT-$\chi$ scheme~\cite{acotnn}, which is designed to
approximately account for production threshold kinematics.

The other partons are parametrized at an initial scale $Q_{0}=1.295$ GeV with adjustable shape parameters.
The parametrizations are essentially like CT10,
but with some minor changes especially for the gluon PDF.
The values of the shape parameters are varied to find the best fit
to the global data set, which we shall describe in Section~\ref{sec:DATA}.
This best fit is obtained by minimizing a global $\chi^{2}$ function
with respect to the input shape parameters,
for different choices for IC.
The fitting procedures include the treatment of systematic
errors and other techniques that have been described in recent
reports on CTEQ-TEA global analysis~\cite{ct10nn}.

We note here that the scale parameters $Q_0$ and $Q_c$, as well as the mass scale introduced in the S-ACOT-$\chi$
scheme could affect the  determination of IC.  In principle, any variation of the scale $Q_0$ can be absorbed
into the parametrization of the initial PDFs, but in practice it can affect the overall $\chi^2$ minimum.
In the absence of IC, any variation of $Q_c$ away from $m_c$
would produce a change in the predictions at the next higher order in $\alpha_s$.
Similar dependence on the  S-ACOT-$\chi$ rescaling variable should occur at higher order in $\alpha_s$ in the
absence of IC. The effects induced from these latter two scales, 
$Q_c$ and the rescaling variable $\chi$, do not cancel in the presence of
nonzero IC, but instead can be considered as part of the defining parametrization of IC.
However, since we are concerned here with the {\em relative} change in the global $\chi^2$ as IC is turned on, we
can consider the dependence on these three auxiliary scales as part of the systematic theoretical uncertainty of the global
analysis, which is no worse than their contribution to the uncertainty in global fits without IC.
Thus, we keep them fixed, while varying the amount of IC. We keep fixed $Q_0=1.295$ GeV, $Q_c=m_c$, and we
use the default definition of the ACOT-$\chi$ variable,
\begin{equation}
\chi = (1+4 m_{c}^{2}/Q^{2})~x,
\end{equation}
for evaluating the coefficient functions in charm production; note that $\chi\sim x$ for $Q\gg m_c$.  Different choices for
these variables may affect the precise limits that we can place on IC, but we do not expect them to change our overall
conclusions.

In contrast to the above three auxiliary scale parameters, we distinguish the charm quark mass parameter of QCD, $m_c$, which enters in the hard matrix elements through the coefficient functions.  Since this is a fundamental parameter
of the theory, it is possible that there is a correlation between the value of $m_c$ and the amount of
intrinsic charm that is physically significant.
We shall postpone the discussion of this further until
Section~\ref{sec:McDependence}.
In this section and the following two sections
we keep the charm mass fixed at the
CTEQ standard value of $m_c=1.3$ GeV.

To examine the dependence on the type and amount of intrinsic charm
we carry out a series of fits, varying the parameter $A$ in
(\ref{eq:modelB}) or (\ref{eq:modelS}).
That is, we minimize $\chi^{2}$ with respect to all variations
of the input parameters, constrained by a fixed value of
the intrinsic charm content, $\hat{c}(x)$, which we specify by its
momentum fraction\footnote{%
The notation $\langle{x}\rangle$ refers here to the momentum fraction
or first moment of the given PDF; it
does \emph{not} signify the
mean value of $x$, which is undefined, in general, if the zero-th moment
is undefined.}
at the scale $Q_c$,
\begin{equation}
\langle{x}\rangle_{\rm IC}
= \int_{0}^{1} x \left[2\hat{c}(x) \right]dx,
\end{equation}
which in turn is determined by $A$.
Here we have multiplied by a factor of 2 in order to include both the $c$ and $\bar{c}$ contributions.
To satisfy the proton momentum sum rule at $Q\ge Q_c$, this momentum fraction has been subtracted from the
total momentum fraction available to the light partons for $Q<Q_c$.

\begin{figure}[H]
\begin{center}
\includegraphics[width=0.8\textwidth]{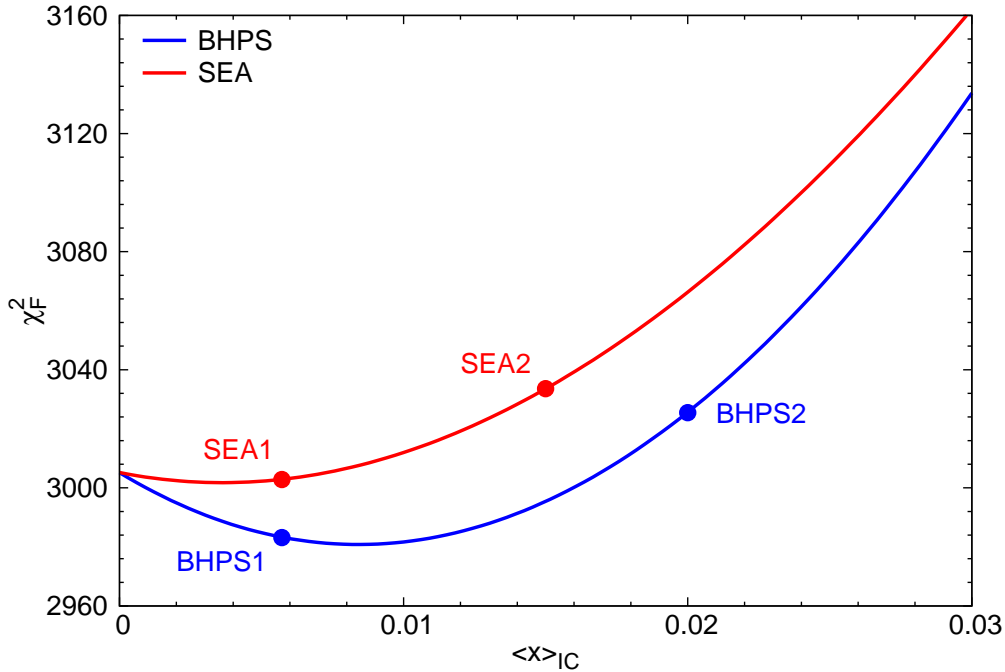}
\end{center}
\caption{The global chi-square function
$\chi_{F}^{2}$ versus charm momentum fraction, $\langle{x}\rangle_{\rm IC}$.
The two approximately parabolic curves are determined
from fits with many values of $\langle{x}\rangle_{\rm IC}$.
Two exemplary fits for each IC model are shown as dots.
Blue: BHPS model;
the dots have $\langle{x}\rangle_{\rm IC}=0.57 \% $ and $2\%$,
which are denoted BHPS1 and BHPS2 below.
Red: SEA model;
the dots have $\langle{x}\rangle_{\rm IC}=0.57 \% $ and $1.5\%$,
which are denoted SEA1 and SEA2 below.
\label{fig:chisqVxic}}
\end{figure}

Figure \ref{fig:chisqVxic} shows the results of the two IC series.
We plot here the global chi-square function, $\chi_{F}^{2}$,
versus the intrinsic charm content, $\langle{x}\rangle_{\rm IC}$,
for the two models of intrinsic charm. The  BHPS model is shown in blue; The SEA model is shown in red.
The function $\chi_F^2$, which is called $\chi^2_{\rm global}$ in Ref.~\cite{ct10nn},
includes the treatment of correlated systematic errors.
The parabolic curves are determined from fits
with many values of $\langle{x}\rangle_{\rm IC}$.
Two exemplary fits for each model are shown as dots.
They have $\langle{x}\rangle_{\rm IC} = 0.57 \%$ and $2 \%$ for the  BHPS models;
and $0.57 \%$ and $1.5 \%$ for the SEA models.
These 4 examples will be used in the subsequent discussion.

Figure~\ref{fig:chisqVxic}  provides a first step toward setting
upper limits on the intrinsic charm.
As $\chi_{F}^{2}$ increases,
the goodness-of-fit to the global data set decreases,
and at some point we judge that the data has ruled out the theory;
this point could define a maximum acceptable value
of $\langle{x}\rangle_{\rm IC}$.
However, we know from experience that relying only on
the value of $\chi_{F}^{2}$ is not always the best measure
of the goodness-of-fit.
For example, there may be PDFs with small \emph{global} $\chi^{2}_{F}$,
but for which one or a few individual data sets are very poorly fit,
balanced by especially good fits to other data sets.
Furthermore, in the global $\chi^2_F$,
a particular experiment that is very sensitive to a fit parameter,
but has a small number of data points will be relatively underweighted compared
to other data sets that are less sensitive, but which have more data points.
Thus, in order that the PDFs
agree to a reasonable degree with all the individual data sets,
we must also look at the $\chi^{2}$ values from
individual experiments,
not just the overall sum of the $\chi^{2}$ values.

To obtain a measure of the goodness-of-fit that includes
the separate values of individual experiments,
we introduce a ``Tier-2 penalty'' $T_{2}(i)$ for each experiment $i$.
This measures the goodness-of-fit for that experiment;
a large value means that the experiment $i$ is not consistent
with the theory.
($T_{2}(i)$ is not simply $\chi_{i}^{2}$,
but they are related.
$T_{2}(i)$ is designed to increase more rapidly than
$\chi_{i}^{2}$ when $\chi_{i}^{2}$ moves beyond the 90\% confidence level.
Additional details are given in the Appendix.)

The dotted curves in Figure \ref{fig:chisq+T2Vxic} show $\chi^{2}_{F}+T_{2}$
versus $\langle{x}\rangle_{\rm IC}$ for the two models of IC.
Our usual choice for the ``tolerance'' of a PDF fit is
\begin{equation}
\Delta\left(\chi_{F}^{2}+T_{2}\right)
= \chi_{F}^{2}+T_{2} - \left(\chi_{F}^{2}+T_{2}\right)_{\rm min}
< 100.
\end{equation}
That is, a set of PDFs with $\Delta\left(\chi_{F}^{2}+T_{2}\right)>100$
is deemed to be such a poor fit to the data that it is ruled out
(at the 90\% confidence level).
From Fig.~ \ref{fig:chisq+T2Vxic} we see that in the present study,
the $T_2$ contribution turns on and rises very sharply with
$\langle{x}\rangle_{\rm IC}$,
thus determining the upper limits on IC.
The sharp increase in slope arises when one
or more experiment becomes poorly fit as the charm momentum
fraction increases.
As we shall see and discuss in more detail in section \ref{sec:DATA},
the dominant constraint on the SEA model is from
the combined HERA charm production measurements;
however, the main constraints on the BHPS model come from
several measurements, but not the HERA charm production
experiments.
We conclude that the upper limits
on $\langle{x}\rangle_{\rm IC}$, at the 90\% C.L., are:
\begin{eqnarray}
\langle{x}\rangle_{\rm IC} &\lesssim & 0.025 {\rm ~~~for~the~BHPS~model}, \nonumber \\
\langle{x}\rangle_{\rm IC} &\lesssim & 0.015 {\rm ~~~for~the~SEA~model}. \nonumber
\end{eqnarray}
Note that the four example PDFs
are all within these limits.
The global data does not rule out BHPS2 (which has $\langle{x}\rangle_{\rm IC}=2\% $),
nor SEA2 (which has $\langle{x}\rangle_{\rm IC}=1.5\% $).
However, these are close to the upper limits for
$\langle{x}\rangle_{\rm IC}$, and can be taken as
representative PDF models with amounts of IC near the largest acceptable value.

\begin{figure}[H]
\begin{center}
\includegraphics[width=0.8\textwidth]{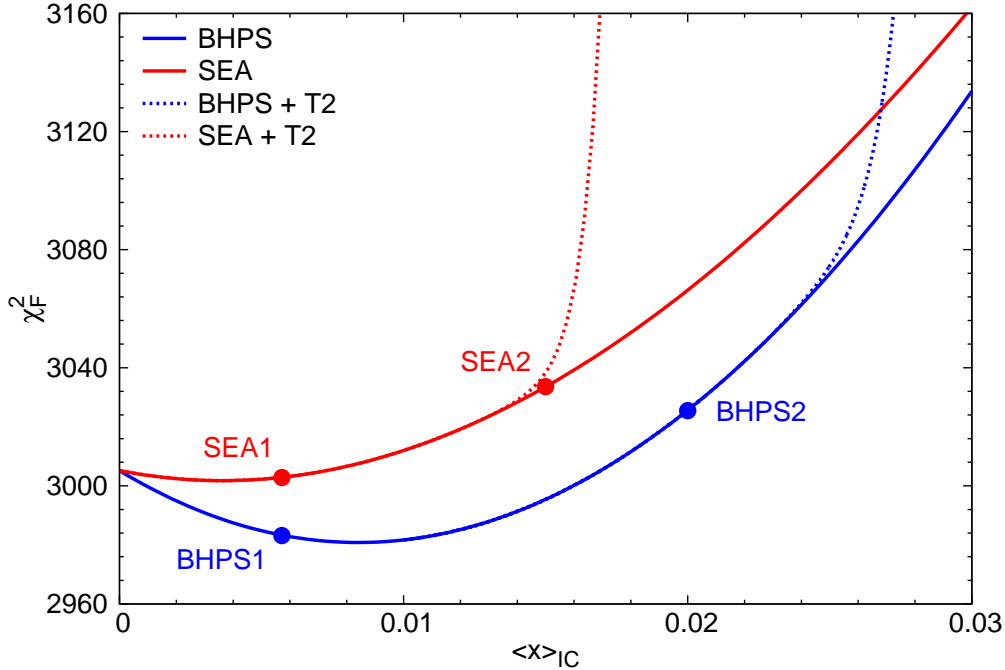}
\end{center}
\caption{%
Similar to Fig. 1, but the
$\chi_{F}^{2}+T_{2}$ versus charm momentum fraction, $\langle{x}\rangle_{\rm IC}$,
is also shown as the dotted curve for both the SEA and BPHS models.
\label{fig:chisq+T2Vxic}}
\end{figure}

Figures~\ref{fig:chisqVxic} and \ref{fig:chisq+T2Vxic} 
show that $\langle{x}\rangle_{\rm IC} > 0.015$
is disfavored for the SEA model,
while the BHPS model allows a larger charm content,
up to about $\langle{x}\rangle_{\rm IC} = 0.025$.
Furthermore, the BHPS model provides a value of
$\chi^{2}_F+T_2$ that is \emph{lower} than that of the standard
CT10 PDFs for $\langle{x}\rangle_{\rm IC} < 0.018$.
The BHPS model with $\langle{x}\rangle_{\rm IC} = 0.009$
gives the best fit to the global data set.
However, the small decrease  $\Delta(\chi_{F}^2+T_{2}) \simeq -20$
is really not significant enough to herald the
discovery of intrinsic charm.
In the rest of the paper we shall  focus on the four IC models
marked as dots in Figs. 1 and 2, labelled as SEA1, SEA2,
BHPS1 and BHPS2. The fraction of intrinsic charm
component for each model is 0.57\%, 1.5\%, 0.57\% and 2\%,
respectively, cf. Table 2.

The result in Fig.~\ref{fig:chisqVxic} is rather different
from the corresponding result from our earlier NLO CTEQ6.5 study~\cite{plt}.
The order of magnitude of the dependence of $\chi_{F}^{2}$
on $\langle{x}\rangle_{\rm IC}$
is comparable to the earlier study.
However, the behaviors of the BHPS model and the SEA model are reversed.
In the current NNLO study,
we find $\chi^{2}_{\rm BHPS} < \chi^{2}_{\rm SEA}$,
leading to a larger upper limit on
$\langle{x}\rangle_{\rm IC}$ for the BHPS model.
In the older NLO study the inequality was the reverse.
Because of the advances in both theory and data, listed above,
we trust that the current results are more realistic regarding
the issue of intrinsic charm.

More extensive comparisons between theory and experiment
are given in Section 3, while a discussion of the correlation of these fits with
the value of the charm mass is deferred to Section 5.

\subsection{The charm quark PDF}

\begin{figure}[H]
\begin{center}
\includegraphics[width=0.4\textwidth]{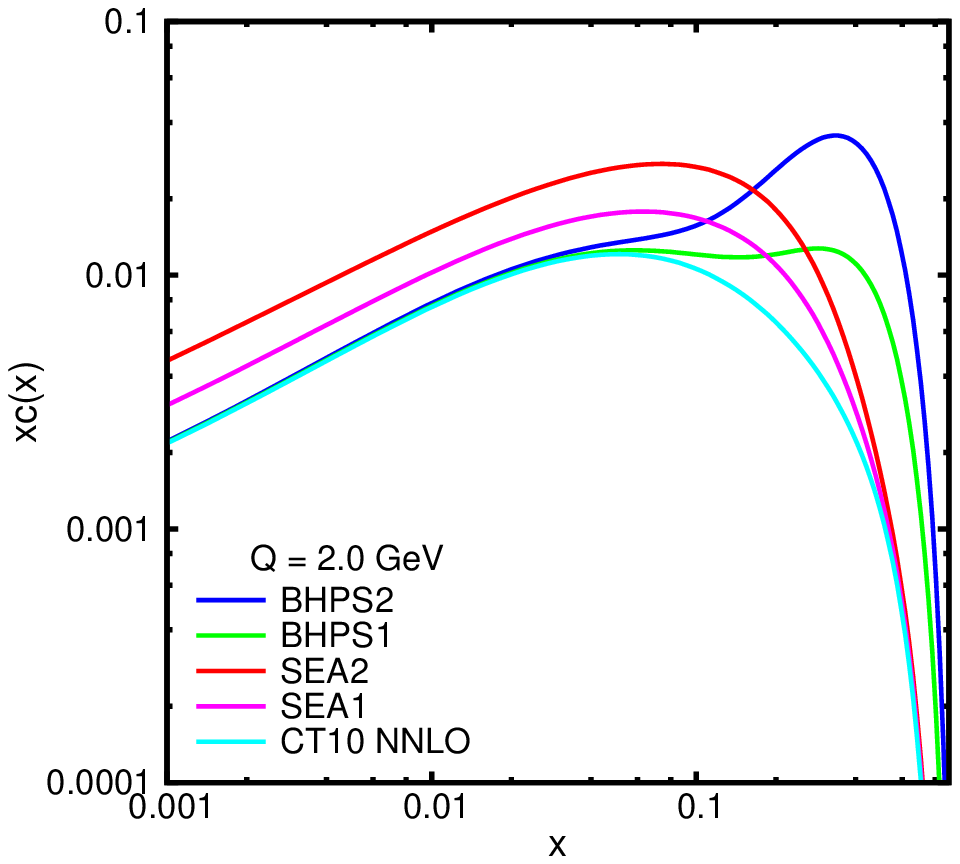}\hspace{.2in}
\includegraphics[width=0.4\textwidth]{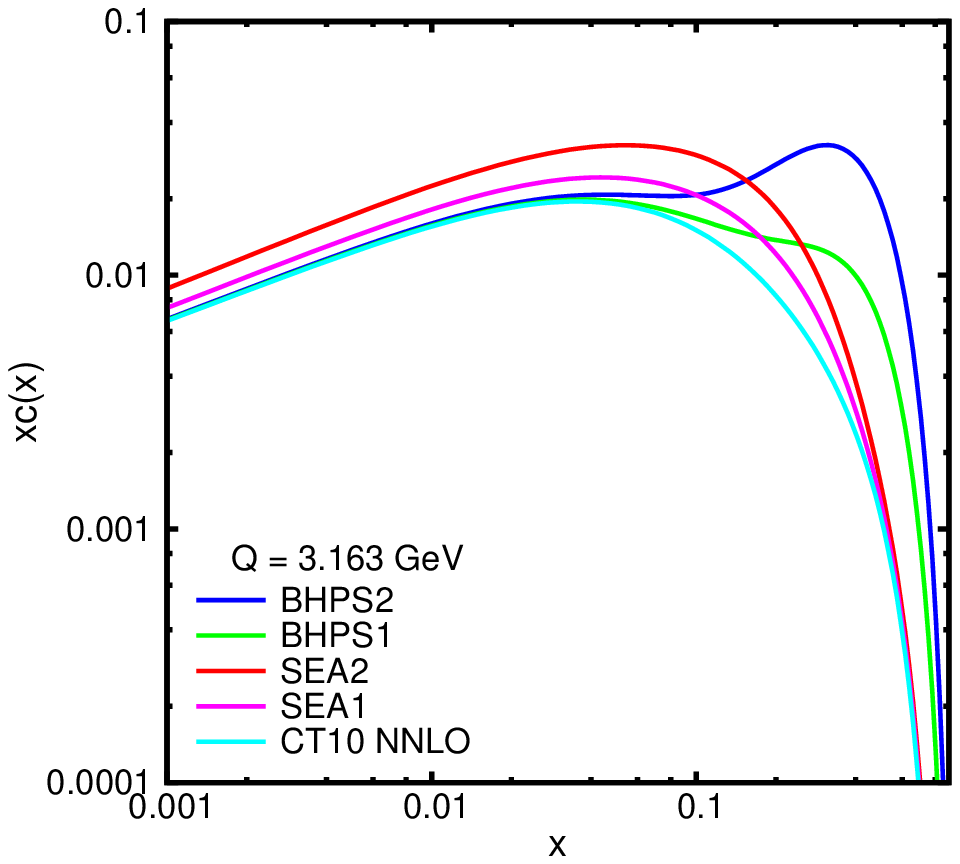}
\includegraphics[width=0.4\textwidth]{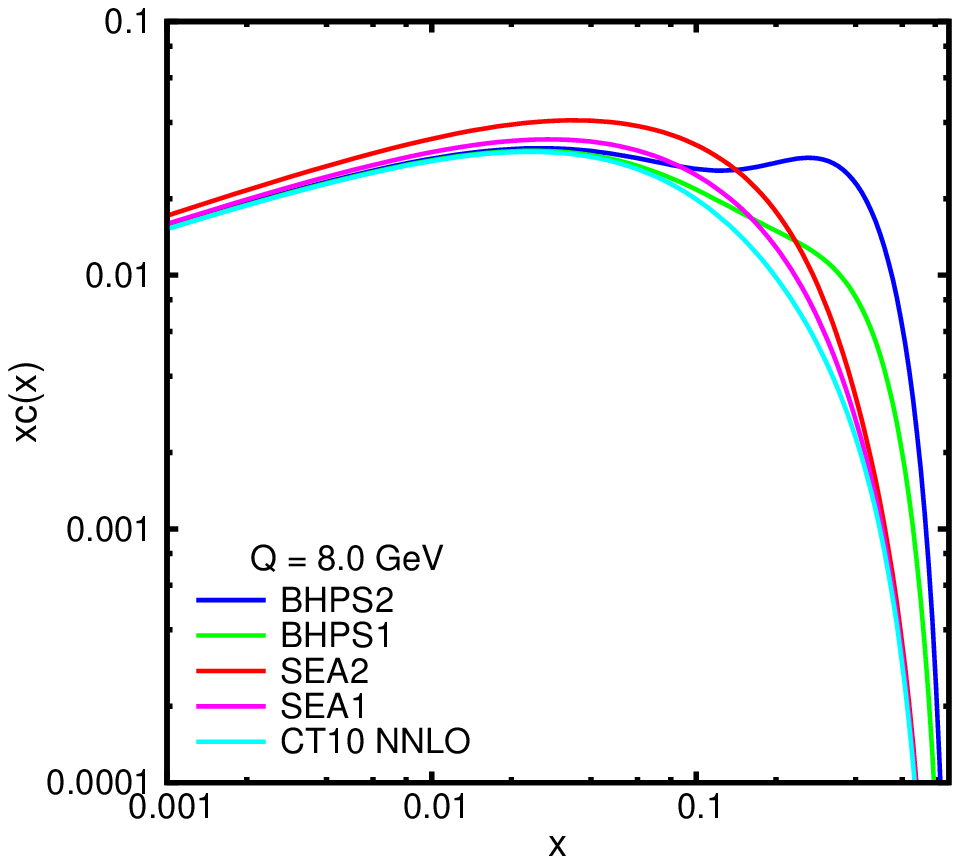}\hspace{.2in}
\includegraphics[width=0.4\textwidth]{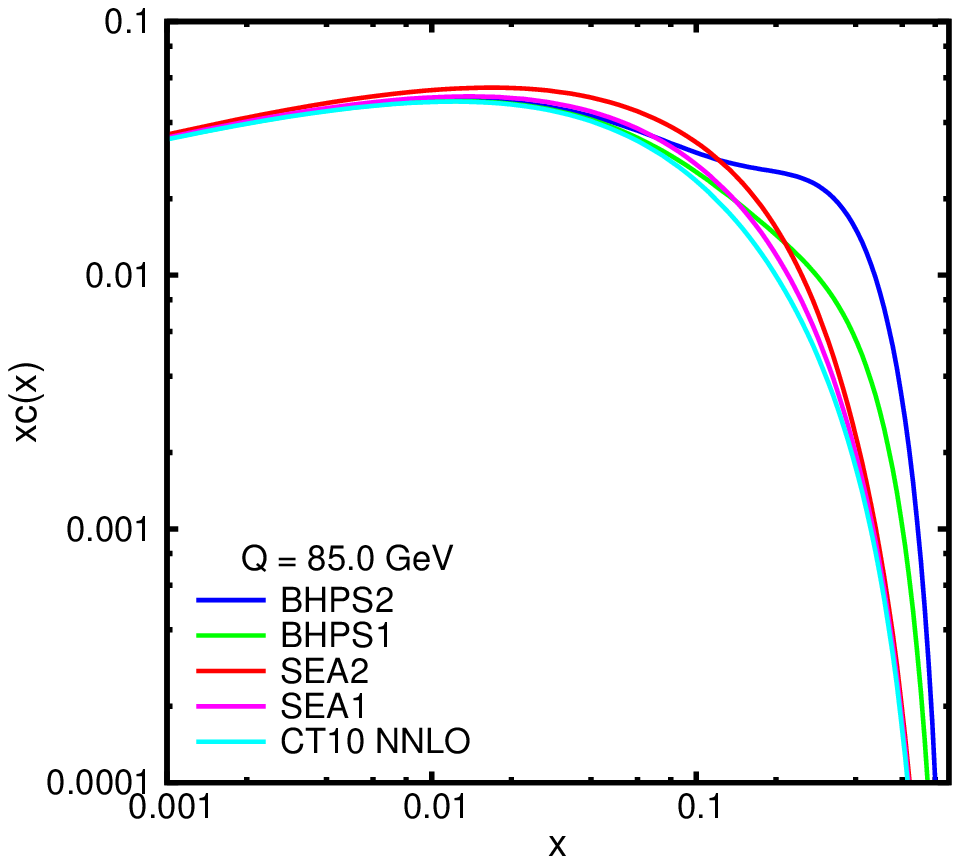}
\end{center}
\caption{Charm quark distribution $x\,c(x,Q)$ from the BHPS1 and BHPS2 PDFs
(which have $0.57\%$ and $2\%$ $\langle{x}\rangle_{\rm IC}$);
from SEA1 and SEA2 PDFs
(which have $0.57\%$ and $1.5\%$ $\langle{x}\rangle_{\rm IC}$);
and from CT10.
The four graphs correspond to $Q = 2.0, 3.16, 8.0, {\rm ~ and ~} 85$ GeV.
\label{fig:cpdf}}
\end{figure}

Figure \ref{fig:cpdf} shows the charm quark PDF for five cases.
In each panel, the function $c(x,Q)$ is shown for the 
BHPS1, BHPS2, SEA1, SEA2 and CT10 PDFs.
The four panels correspond to four values of $Q$:
$Q = 2.0, 3.16, 8.0, 85\,{\rm GeV}$.
CT10 has no IC; $\hat{c}(x)=0$.
But the charm distribution evolves rapidly from $Q_{c}$ to 2 GeV.
For $Q > 8$ GeV and $x  < 0.01$,
the CT10 charm PDF is comparable to the IC models.

The BHPS model has a ``valence-like'' charm distribution at $Q_{c}$.
So for small $Q$, the charm density with $x > 0.1$ is significantly larger
than our standard CT10 charm PDF.
Even for $Q$ as large as 85 GeV,
$c(x,Q)$ is still notably larger for the BHPS model than for CT10, for large $x$.
On the other hand, for small $x$, say $x < 0.05$,
the BHPS model and CT10 charm densities are approximately equal
for $Q > 8$ GeV.
Because the IC in this model is concentrated at large $x$,
the resulting PDFs can fit most data comparably well as CT10,
as long as $\langle{x}\rangle_{\rm IC}$ is not too large;
cf.\ Section 3.

The SEA model has a ``sea-like'' charm distribution at $Q_{c}$.
There is a large $c$ density at $Q_{c}$
for small and intermediate $x$ ranges, say $x < 0.2$.
The charm PDF evolves rapidly from $Q_{c}$ to 2 GeV,
but nevertheless $c(x,Q)$ continues to be notably larger
than CT10 at small $x$, even for $Q$ as large as 8 GeV.
Because of this large charm density at low $x$,
the SEA model has a more difficult time fitting the combined HERA charm data
 than CT10, or than the BPHS model 
for equal values of  $\langle{x}\rangle_{\rm IC}$; cf.\ Section 3.

\begin{figure}[H]
\begin{center}
\includegraphics[width=0.8\textwidth]{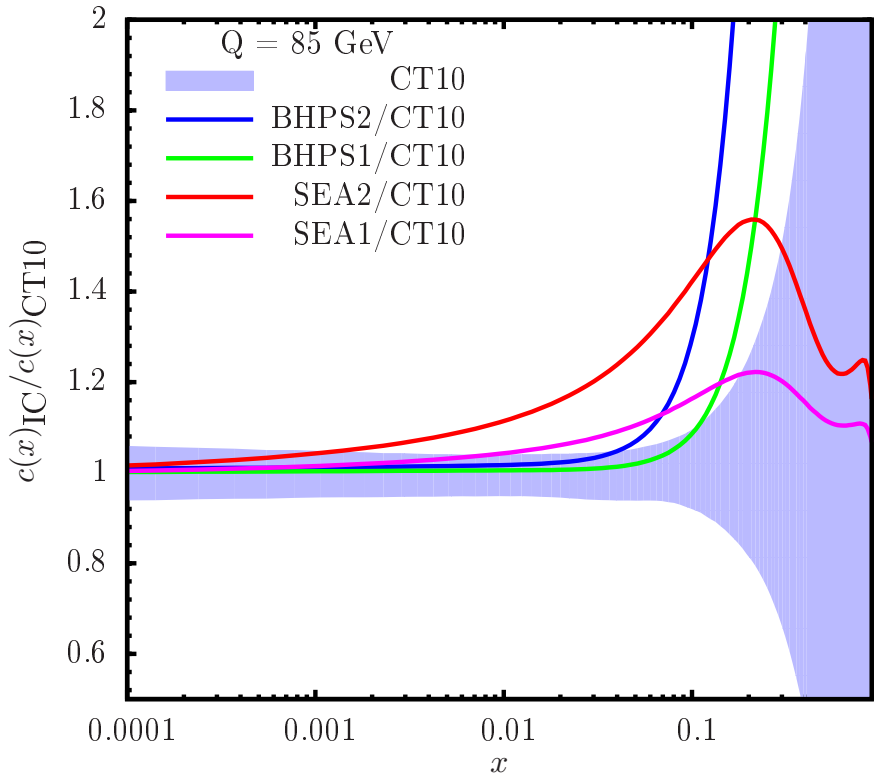}
\end{center}
\caption{
Ratios of $c(x,Q)_{\rm IC}/c(x,Q)_{\rm CT10}$ for $Q = 85\ {\rm GeV}$.
\label{fig:cratios}}
\end{figure}

To emphasize the differences between the models,
relevant to LHC experiments,
Figure \ref{fig:cratios} compares $c(x,Q)$
for the four examples of the BHPS and SEA models,
at the large momentum scale $Q = 85\,{\rm GeV}$.
For each case the ratio $c(x,Q)_{\rm IC}/c(x,Q)_{\rm CT10}$
is plotted as a function of $x$.
The shaded region is the uncertainty band for
the charm distribution of CT10 at $Q=85\ {\rm GeV}$.
Note that the charm distributions with IC are allowed to be outside
the uncertainty band of CT10, since CT10 was created with the
constraint that the charm PDF was radiatively generated.

Putting intrinsic charm into the $c$ and $\overline{c}$
distributions will, of course, directly affect predictions involving
charm quarks in the initial state.
But there will also be indirect effects.
Because of the momentum sum rule, other parton distributions
must change to balance the change in the charm distribution.
Therefore we should also look at other parton distributions
in the PDF sets with intrinsic charm.

\begin{figure}[H]
\begin{center}
\includegraphics[width=0.8\textwidth]%
{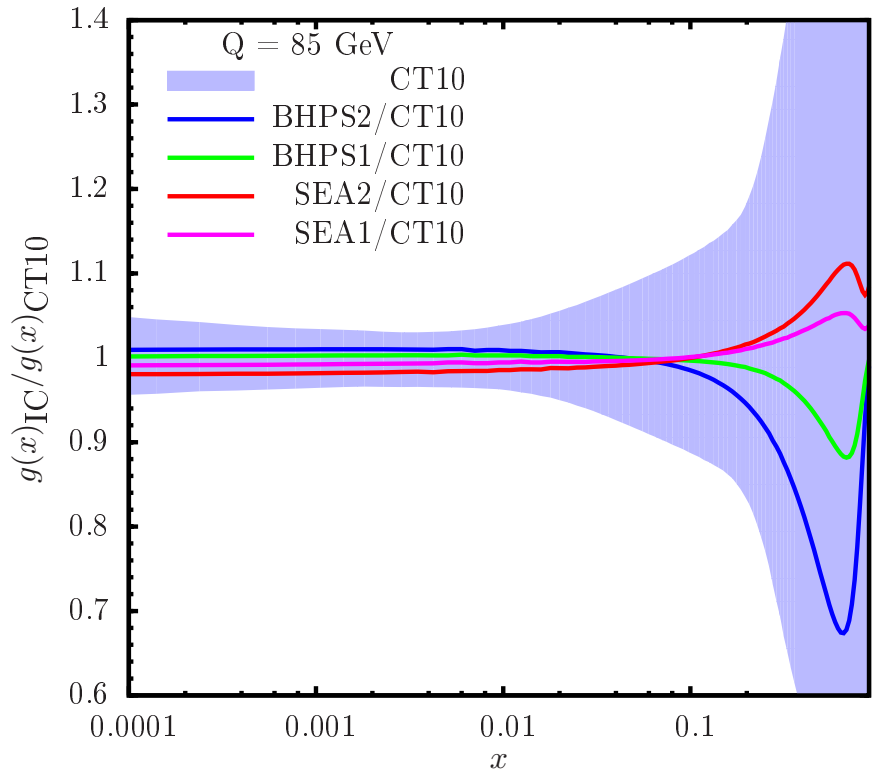}
\end{center}
\caption{
Ratios of $g(x)_{\rm IC}/g(x)_{\rm CT10}$
for $Q = 85\ {\rm GeV}$.
\label{fig:gluratios}}
\end{figure}

\begin{figure}[H]
\begin{center}
\includegraphics[width=0.8\textwidth]%
{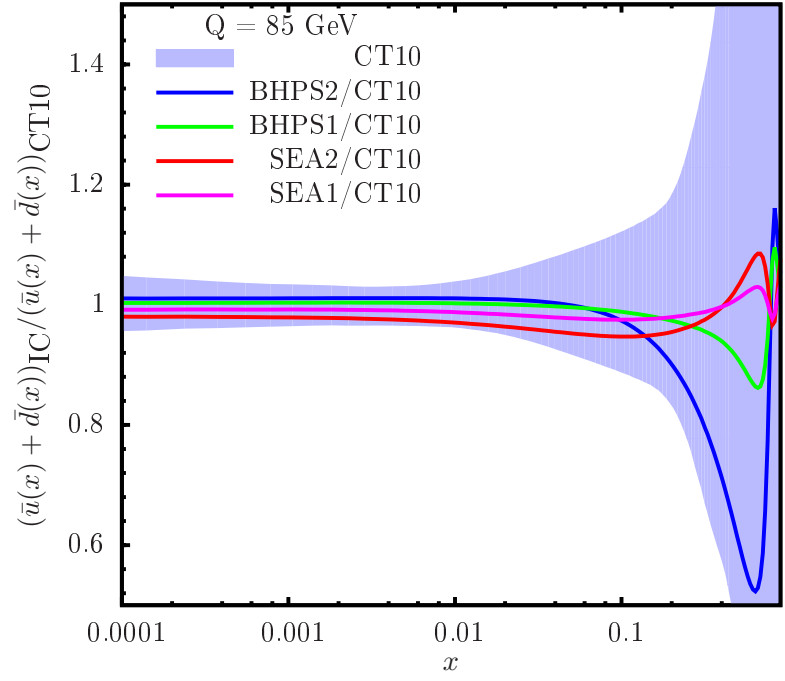}
\end{center}
\caption{
Ratios of $(\bar u(x)+\bar d(x))_{\rm IC}/(\bar u(x) +\bar d(x))_{\rm CT10}$
for $Q = 85\ {\rm GeV}$.
\label{fig:ubrratios}}
\end{figure}

Figure \ref{fig:gluratios} shows the gluon PDF for the
same four examples of the BHPS and SEA models as in Fig.\ \ref{fig:cratios},
again plotting the ratio to CT10,
for the momentum scale $Q = 85\ {\rm GeV}$.
Figure \ref{fig:ubrratios} shows the same
for the $\bar u + \bar d$ quark PDF.
The shaded regions show the uncertainty bands for CT10.
We note that for the BHPS model,
the presence of intrinsic charm pulls momentum from the gluon
and the $\bar u + \bar d$ quark at large $x$;
whereas for the SEA model, the presence of IC pulls
momentum from the $\bar u + \bar d$ quark at small $x$.
This will affect LHC predictions in interesting ways;
cf.\ Section 4.
Before concluding this section, we note that
each IC fit is a central fit,
with a different assumption on the intrinsic charm distribution
at the scale $Q_c$.  Hence, if we would compute eigenvector
(Hessian) uncertainties for these IC fits, we would obtain an
uncertainty range that expands on the CT10 uncertainty range
for these gluon and quark distributions.

\clearpage

\section{Experimental data sets and intrinsic charm
\label{sec:DATA}}

The experimental data sets included in the global analysis that led to the
results of Section~\ref{sec:RESULTS} are listed in Table~\ref{tab:EXP_bin_ID}.
With a few exceptions, they are the same as the data used in the standard
CT10 fit~\cite{ct10nn}.
We note that the CT10 PDFs were fitted to the individual $F_2^{c}$
data sets from both H1 and ZEUS Collaborations. As shown in Table 1,
in this study we have replaced those individual $F_2^{c}$ data sets
by the {\it combined} H1 and ZEUS reduced cross section data
(called Data Set 147) for charm production at HERA~\cite{HERAX0c}.

The impact of an intrinsic charm component on the fit to the different data sets can
be visualized in several different ways.  In Table~\ref{tab:EXP_bin_ID}, we have
listed the number of data points, $N_{pt}$, for each experiment, as well as the
$\chi^{2}/N_{pt}$ for the CT10 fit, the BHPS2 fit, and the SEA2 fit,
which have charm momentum fractions ($\langle{x}\rangle_{\rm IC}$)
of 2\% and 1.5\%, respectively.
The change in $\chi^{2}/N_{pt}$ in the last two columns from the CT10 fit
indicates which data sets
are in conflict with a large intrinsic charm content.

\def \Dzero {D0}

{
\begin{table}[p]
\begin{tabular}{|l|l|l|l|l|l|}
\hline
\textbf{ID } & \textbf{Experimental data set}
& $N_{pt}$ & CT10 & BHPS(0.020) & SEA(0.015) \tabularnewline
\hline
\hline
101  & BCDMS $F_{2}^{p}$~\cite{Benvenuti:1989rh}
& 339  & 1.158 & 1.087 & 1.220 \tabularnewline
\hline
102  & BCDMS $F_{2}^{d}$~\cite{Benvenuti:1989fm}
& 251  & 1.157 & 1.119 & 1.187 \tabularnewline
\hline
103  & NMC $F_{2}^{p}$~\cite{Arneodo:1996qe}
& 201  & 1.656 & 1.668 & 1.582 \tabularnewline
\hline
104  & NMC $F_{2}^{d}/F_{2}^{p}$~\cite{Arneodo:1996qe}
& 123  & 1.210 & 1.311 & 1.207 \tabularnewline
\hline
108  & CDHSW $F_{2}^{p}$~\cite{Berge:1989hr}
&  85  & 0.832 & 0.833 & 0.781 \tabularnewline
\hline
109  & CDHSW $F_{3}^{p}$~\cite{Berge:1989hr}
&  96  & 0.809 & 0.867 & 0.810 \tabularnewline
\hline
110  & CCFR $F_{2}^{p}$~\cite{Yang:2000ju}
&  69  & 0.989 & 1.105 & 0.943 \tabularnewline
\hline
111  & CCFR $xF_{3}^{p}$~\cite{Seligman:1997mc}
&  86  & 0.387 & 0.416 & 0.417 \tabularnewline
\hline
124  & NuTeV $\nu$ di-$\mu$ SIDIS~\cite{Mason:2006qa}
&  38  & 0.781 & 0.836 & 0.745 \tabularnewline
\hline
125  & NuTeV $\overline{\nu}$ di-$\mu$ SIDIS~\cite{Mason:2006qa}
&  33  & 0.852 & 0.905 & 0.864 \tabularnewline
\hline
126  & CCFR $\nu$ di-$\mu$ SIDIS~\cite{Goncharov:2001qe}
&  40  & 1.195 & 1.204 & 1.145 \tabularnewline
\hline
127  & CCFR $\overline{\nu}$ di-$\mu$ SIDIS~\cite{Goncharov:2001qe}
&  38  & 0.692 & 0.728 & 0.655 \tabularnewline
\hline
147 & HERA charm production~\cite{HERAX0c}
&  47  & 1.187 & 1.185 & 1.424 \tabularnewline
\hline
159  & Combined HERA1 DIS~\cite{:2009wt}
& 579  & 1.068 & 1.086 & 1.059 \tabularnewline
\hline
201  & E605 DY process  $\sigma(pA)$~\cite{Moreno:1990sf}
& 119  & 0.804 & 0.845 & 0.796 \tabularnewline
\hline
203  & E866 DY process  $\sigma(pd)/(2\sigma(pp))$~\cite{Towell:2001nh}
&  15  & 0.658 & 0.789 & 0.718 \tabularnewline
\hline
204  & E866 DY process  $\sigma(pp)$~\cite{Webb:2003ps}
& 184  & 1.271 & 1.286 & 1.277 \tabularnewline
\hline
225  & CDF Run-1 $W$ charge asymmetry~\cite{Abe:1996us}
&  11  & 1.292 & 1.191 & 1.285 \tabularnewline
\hline
227  & CDF Run-2 $W$ charge asymmetry~\cite{Acosta:2005ud}
&  11  & 0.978 & 0.995 & 0.978 \tabularnewline
\hline
231  & \Dzero Run-2 $W$ charge asymmetry~\cite{Abazov:2008qv}
&  12  & 1.928 & 2.006 & 1.972 \tabularnewline
\hline
234  & \Dzero Run-2 $W$ charge asymmetry~\cite{Abazov:2007pm}
&   9  & 1.501 & 1.709 & 1.371 \tabularnewline
\hline
260  & \Dzero Run-2 Z rapidity dist.~\cite{Abazov:2006gs}
&  28  & 0.580 & 0.551 & 0.550 \tabularnewline
\hline
261  & CDF Run-2 Z rapidity dist.~\cite{Aaltonen:2010zza}
&  29  & 1.586 & 1.466 & 1.535 \tabularnewline
\hline
504  & CDF Run-2 inclusive jet~\cite{Aaltonen:2008eq}
&  72  & 1.398 & 1.431 & 1.311 \tabularnewline
\hline
514  & \Dzero Run-2 inclusive jet~\cite{:2008hua}
& 110  & 1.044 & 0.950 & 1.012 \tabularnewline
\hline
\hline
  & \textbf{Totals:} & $2625$  & $3005$  & $3034$ & $3026$ \\
\hline
\end{tabular}
\caption{Experimental data sets employed in the CT10 analysis.
$N_{pt} = $ the number of points in the data set.
The final three columns show $\chi^{2}/N_{pt}$ for each data set for:
CT10,
the BHPS model with $\langle{x}\rangle_{\rm IC} = 0.020$,
and the SEA model with $\langle{x}\rangle_{\rm IC} = 0.020$.
The last row shows the global $\chi^{2}_{F}$ for each PDF set 
for a total of 2625 data points.
\label{tab:EXP_bin_ID}}
\end{table}
}

One especially interesting data set for this paper is the \emph{combined}
H1 and ZEUS data for charm production in deep-inelastic $ep$ collisions
at HERA~\cite{HERAX0c}.
This is Data Set 147 in Table \ref{tab:EXP_bin_ID}.
Charmed particles can be produced in two ways in an $ep$ collision:
the charm-excitation process $\gamma + c \rightarrow c + X$,
 and the charm-creation process $\gamma + g \rightarrow c+\overline{c}$.
 These are combined consistently in the calculations using the prescription
 introduced by ACOT~\cite{acot}.  However, in the presence of intrinsic charm,
 the former process becomes particularly important, since it is
 directly proportional to the $c$
and $\overline{c}$ component of the proton.  Thus, one would expect data set
147 to be particularly sensitive to intrinsic charm.  In addition, this
set of combined H1 and ZEUS data has smaller systematic errors than
the separate data sets, which were used in our previous analysis~\cite{plt}.
Therefore we are interested to assess the influence of this
newly available high-precision combined data.

\begin{figure}[tbh] \begin{center}
$
\begin{array}{c}
\includegraphics[width=0.8\textwidth]{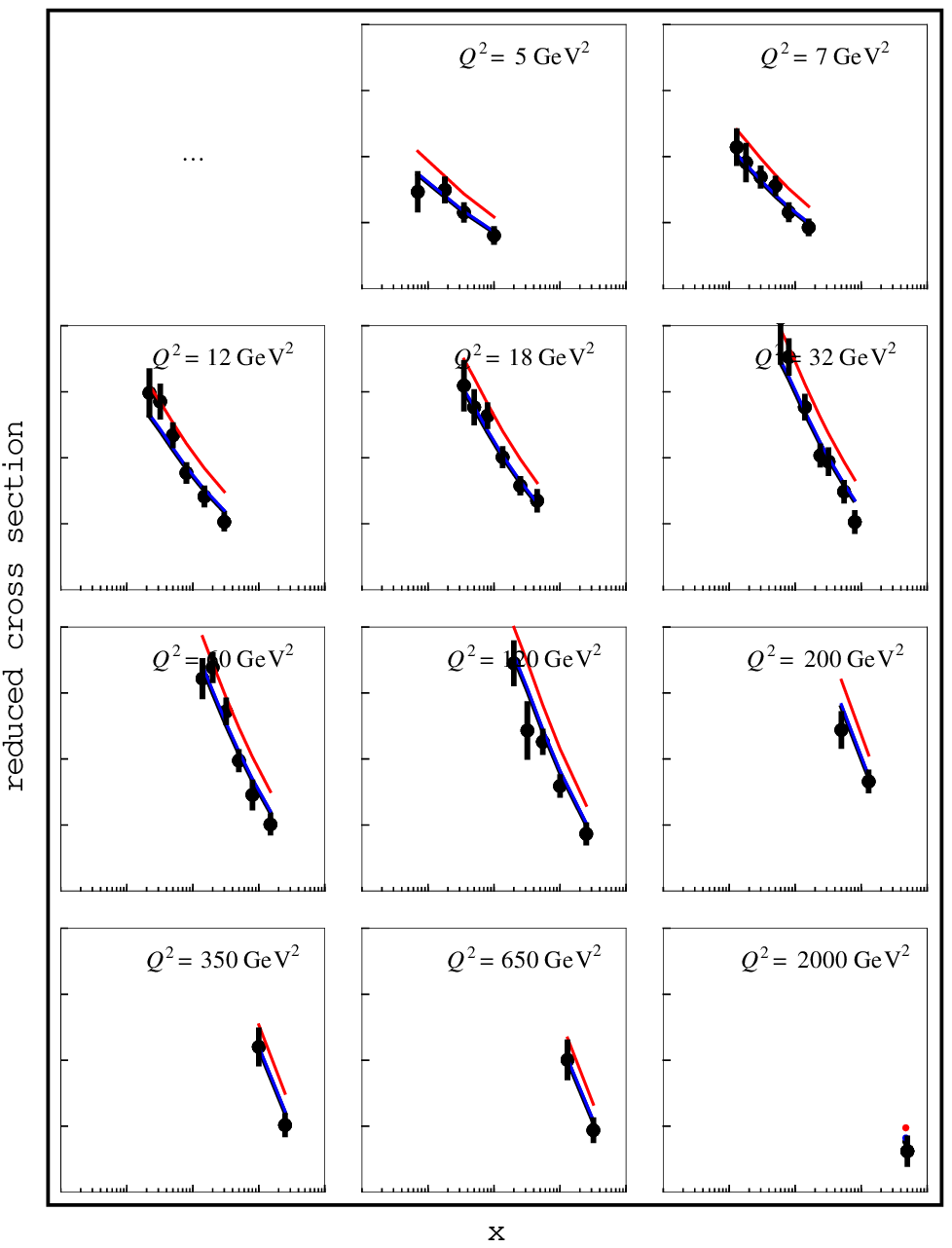}
\end{array}
$
\end{center}
\caption{Comparison of theory and data
for the H1 and ZEUS combined data on charm production
in $ep$ collisions at HERA.
Theory: The blue curves are the result for BHPS2;
the red curves are the result for SEA2.
The data is plotted as points;
the error bars are the total errors.
The CT10 prediction cannot be distinguished from that given
by BHPS2.
\label{fig:HERAc}}
\end{figure}

Figure \ref{fig:HERAc} shows a comparison of the H1 and ZEUS combined
data on charm production with the theory predictions using IC  parton distributions.
The agreement between data and theory for BHPS2 is satisfactory,
and almost the same as for the CT10 PDFs.
On the other hand, we see systematic differences between the data and
SEA2, whose prediction is consistently higher than the data over the
full range of $Q^2$ and $x$.
Table \ref{tab:HERAc} lists the values of $\chi^{2}$ for CT10 and
our four representative IC models with different amounts of IC.
In general the models with sea-like intrinsic charm do worse than the standard
fit or the fits with the BHPS model with valence-like intrinsic charm.
The sea-like models have more charm at low values of $x$,
which is disfavored by this HERA data with small $x$.
We note, however, that the increase in $\chi^2$ for SEA2
is only about 13 units, to be compared to 47 data points.

\begin{table}[p]
\begin{center}
\begin{tabular}{|l|l|l|}
\hline
Model & $\langle{x}\rangle_{\rm IC}$ & $\chi^{2}$\\
\hline
CT10 & 0\% & 55.75 \\
\hline
BHPS1 & 0.57\% & 55.78 \\
BHPS2 & 2\% & 56.29 \\
\hline
SEA1 & 0.57\% & 57.60 \\
SEA2 & 1.5\% & 68.50 \\
\hline
\end{tabular}\end{center}
\caption{$\chi^{2}$ for different models of IC,
for data set 147.
This is the combined H1 and ZEUS data
for inclusive charm production
in $\overline{e}p$ or $ep$ collisions at HERA.
The number of data points is $N_{pt} = 47$.
\label{tab:HERAc}}
\end{table}

The impact of intrinsic charm on the other data sets
in Table~\ref{tab:EXP_bin_ID}
is not so obvious as for the charm production set 147.
Some data sets favor the SEA models,
while other data sets favor the BHPS models,
while still others disfavor both.
One other data set to note in particular is data set 101,
which is the BCDMS $F_2^p$ data.
Although the decrease in $\chi^2/N_{pt}$ for BHPS2,
and the increase in $\chi^2/N_{pt}$ for SEA2,
are not obviously significant,
the large number of data points $N_{pt}=339$ ensures that these
have an affect on the total $\chi^2$.
In this case the BHPS2 fit has $\chi^2$ reduced by 24 units
compared to CT10, while the SEA2 fit has $\chi^2$ increased by
21 units.
The net effect is that the full combination of all data sets
to the total $\chi^2$ disfavors the SEA model of intrinsic charm
more than the BHPS model of intrinsic charm,
for a given intrinsic charm momentum fraction,
as seen in Fig.~\ref{fig:chisqVxic}.

\subsection{Spartyness: an equivalent gaussian variable}

From the discussion in the last paragraph, we can see that the naive use of
total $\chi^2$ as the discriminating variable may overweight data sets
with large numbers of points,
even if the correlation with the fitting parameter is not very significant.
It was for this reason that the Tier-2 penalty was added to $\chi^2$
in the global fitting program.
The Tier-2 penalty makes use of a variable $S_{n}$,
which we call \emph{spartyness},
which gives a measure of the goodness-of-fit
for each of the individual data sets.
It is defined precisely in the Appendix.

In words, we obtain the spartyness for a particular data set by mapping
its $\{\chi^2,N\}$ value, assumed to obey a chi-square probability distribution,
onto the variable $S_{n}$ which has the same probability but for a standard Gaussian
distribution.\ (cf.\ Appendix)

The values of $S_{n}$ can then be interpreted in terms of probabilities
in a normal distribution.
Fits with $S_{n}$ between -1 and 1 are accepted as reasonable,
within the errors.
Fits with $S_{n} > 3$ are considered poor fits.
Fits with $S_{n} < -3$ actually fit the data much better than one would expect
from normal statistical analysis; i.e., they have anomalously small residuals,
presumably because the true experimental errors are smaller than
the published values.

\begin{figure}[tbh]
\begin{center}
$
\begin{array}{c}
\includegraphics[width=0.8\textwidth]{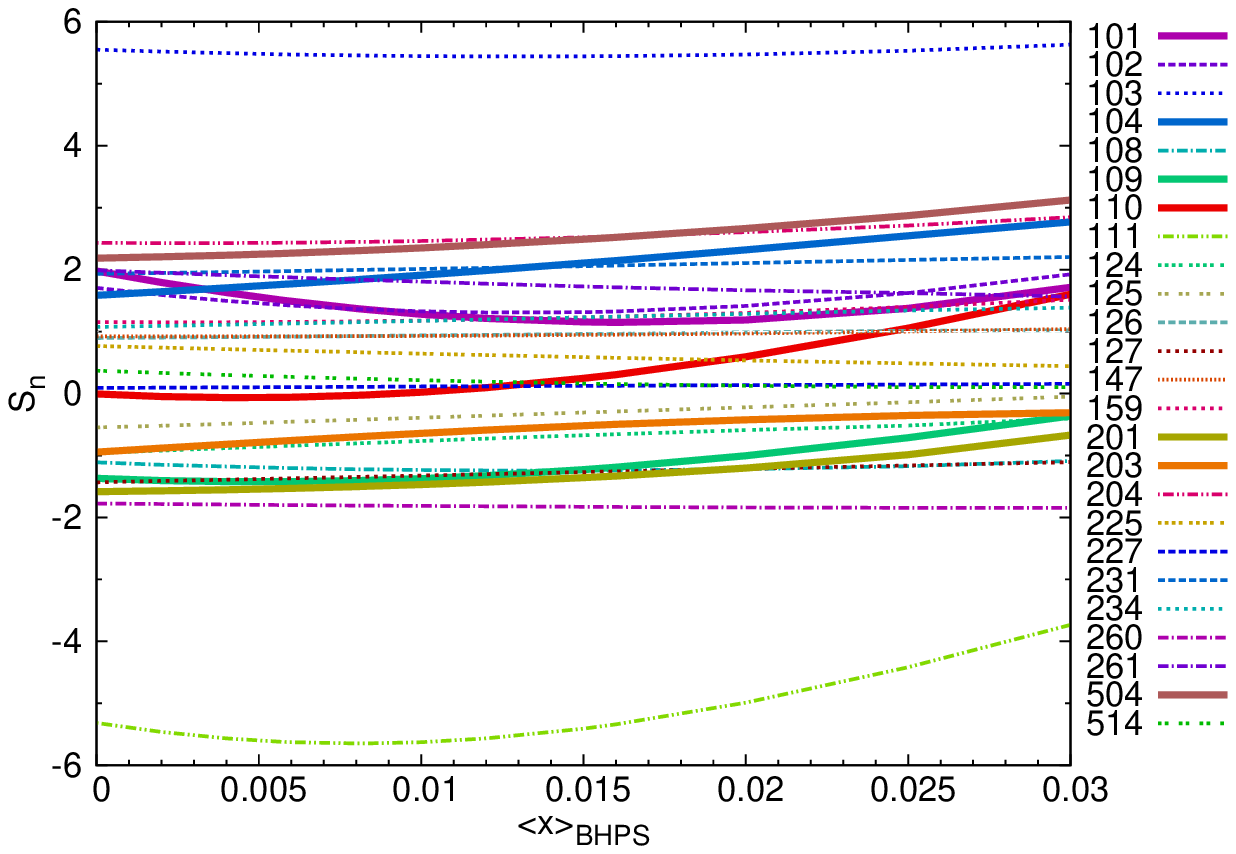}\\
\includegraphics[width=0.8\textwidth]{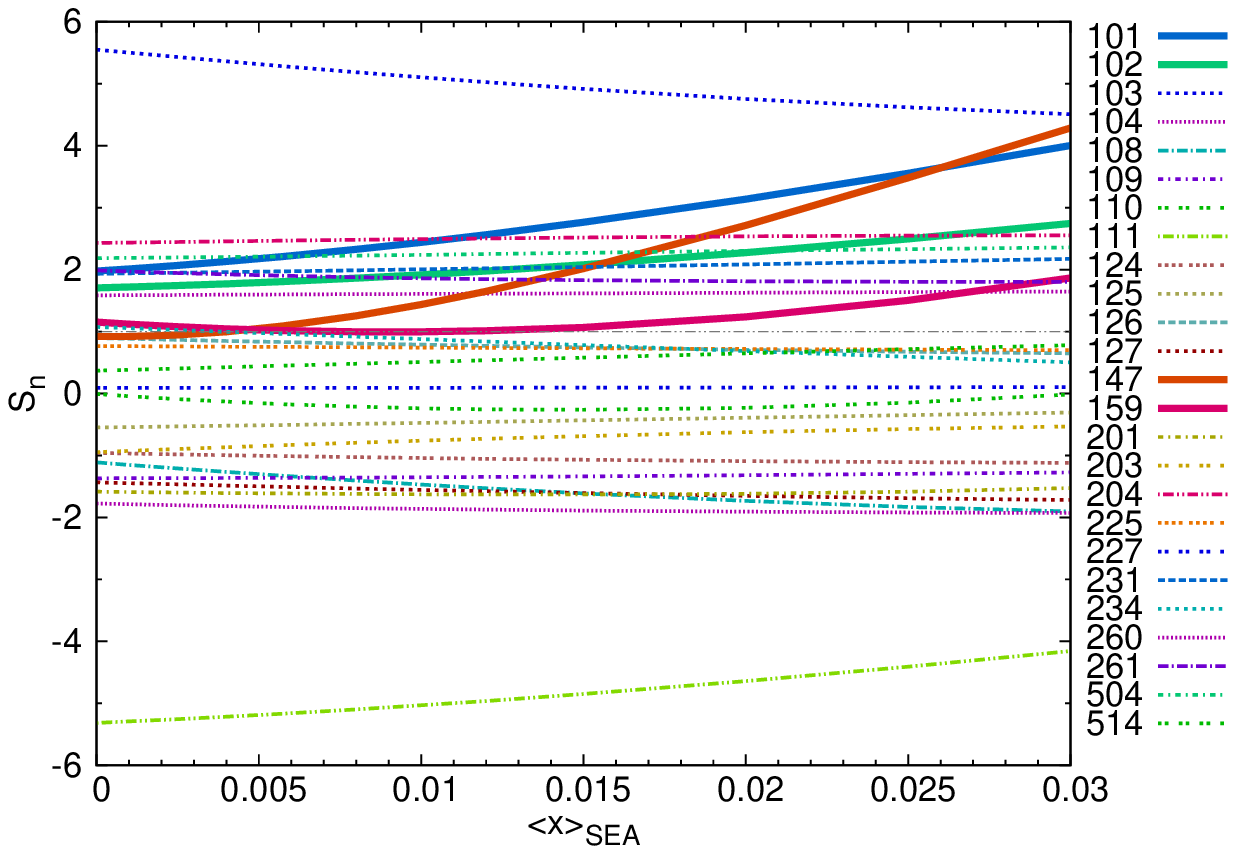}\\
\end{array}
$
\end{center}
\caption{Comparing the ``spartyness'' $S_{n}$ as a function of charm momentum
fraction,
$\langle{x}\rangle_{\rm IC}$, for two models of intrinsic charm.
Upper: BHPS model; lower: SEA model.
\label{fig:spartyall}}
\end{figure}

\begin{figure}[tbh]
\begin{center}
$
\begin{array}{c}
\includegraphics[width=0.8\textwidth]{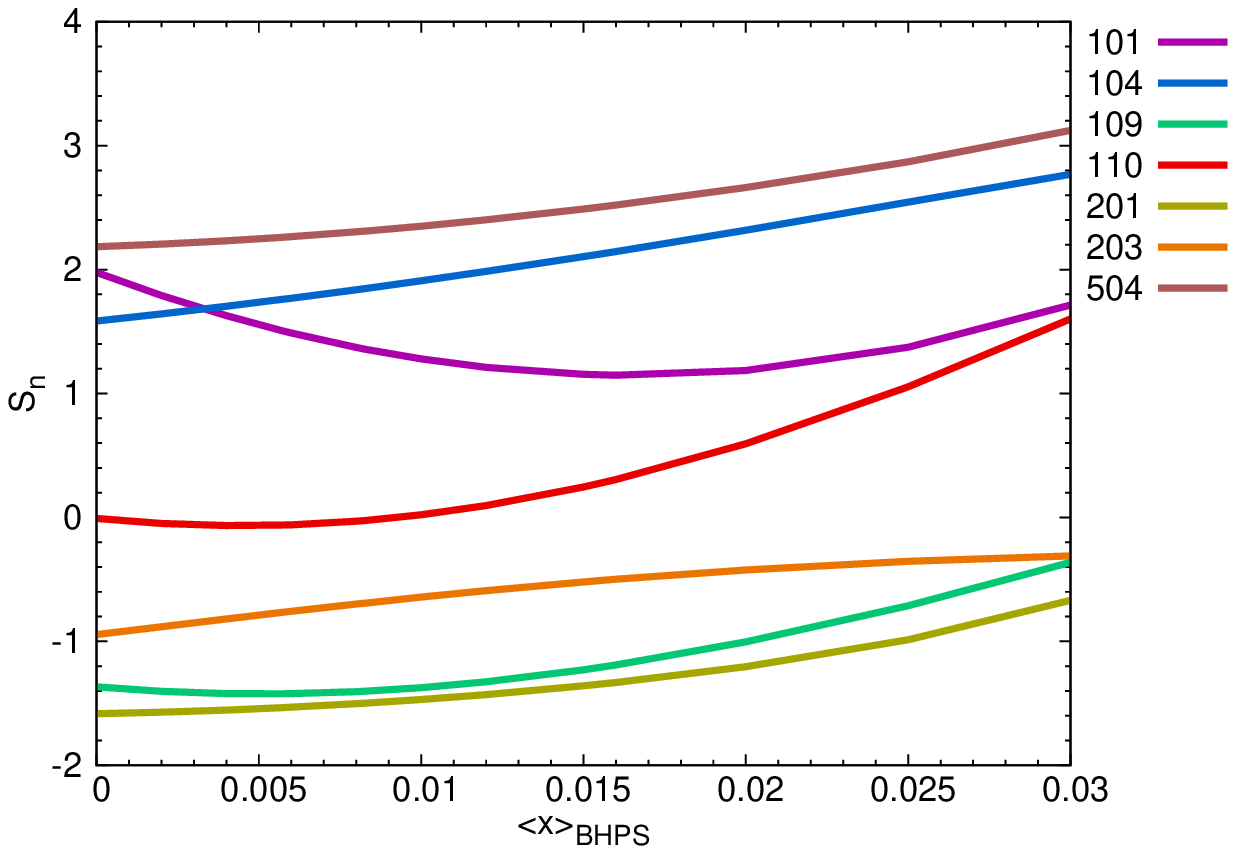}\\
\includegraphics[width=0.8\textwidth]{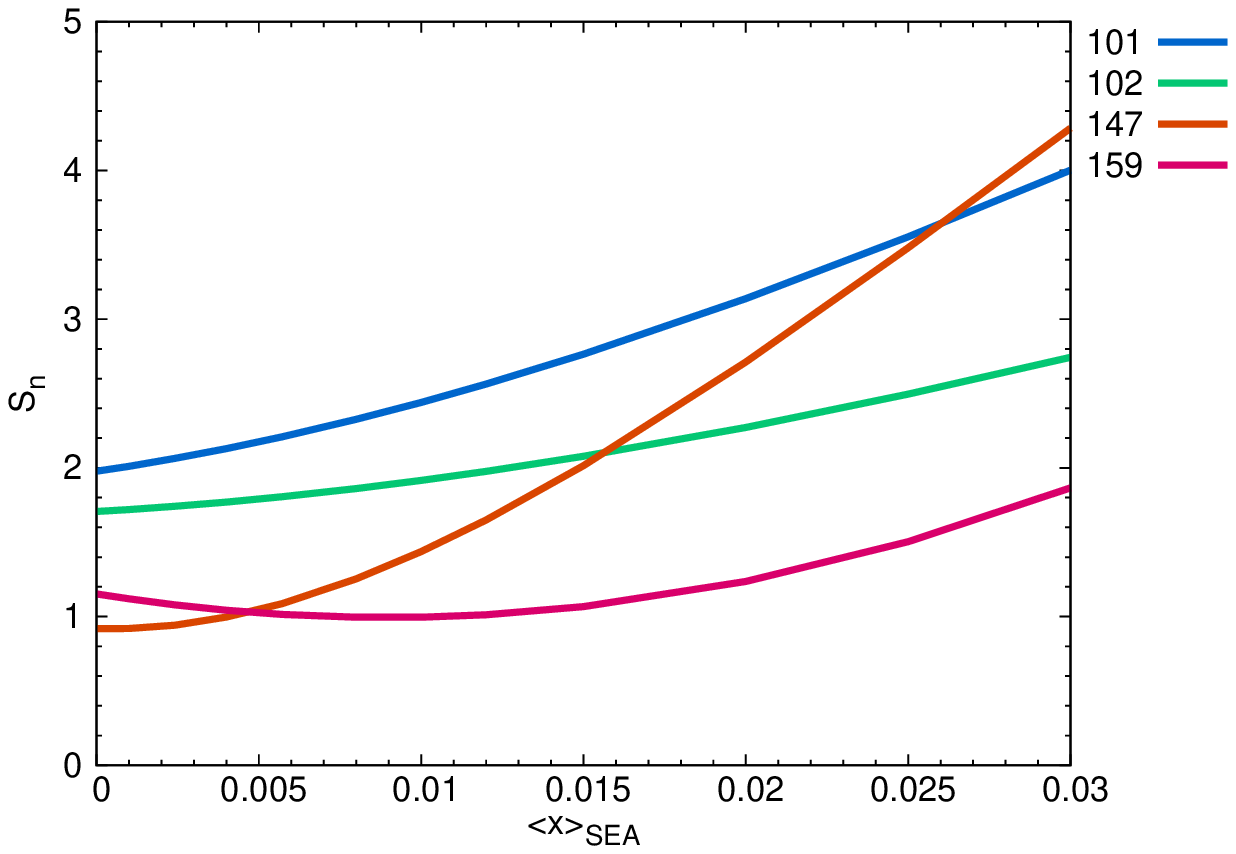}\\
\end{array}
$
\caption{Comparing the ``spartyness'' $S_{n}$ as a function of charm momentum fraction,
$\langle{x}\rangle_{\rm IC}$,
showing the data sets for which $S_{n}$ changes the most significantly.
Upper: BHPS model; lower: SEA model.
\label{fig:spartysig}}
\end{center} \end{figure}

Figure \ref{fig:spartyall} shows $S_{n}$ for all 25 individual data sets,
for the two models of IC, as a function of $\langle{x}\rangle_{\rm IC}$.
The BHPS model is shown above, and The SEA model is shown below.
To focus on the data sets which are most affected
by intrinsic charm, we select out just these in
Figure \ref{fig:spartysig}.

For the BHPS model we see that several of the data sets have a slow
increase in spartyness with momentum fraction $\langle{x}\rangle_{\rm IC}$,
while data set 101 (BCDMS $F_2^p$) has a significant decrease in spartyness
for small $\langle{x}\rangle_{\rm IC}$, and then levels off for larger
$\langle{x}\rangle_{\rm IC}$.  This is consistent with the results
of Fig.~\ref{fig:chisqVxic}
which showed that the BHPS models
for $\langle{x}\rangle_{\rm IC} \lesssim 0.015$
were actually slightly favored over zero intrinsic charm,
but increasing $\langle{x}\rangle_{\rm IC}$
further decreased the overall goodness of fit for the BHPS model.
The increase in $\chi_{F}^{2}$ or $\chi_{F}^{2}+T_{2}$
seen in Figs.\ 1 or 2 at $\langle{x}\rangle_{\rm IC} \lesssim 0.025$
for the BHPS model can be attributed to several experiments:
ID 104 (NMC $F_{2}^{d}/F_{2}^{p}$),
ID 110 (CCFR $F_{2}^{p}$) and
ID 504 (CDF Run-2 inclusive jets);
note that $S_{n}$ increases for these experiments
in Fig.\ \ref{fig:spartysig}.
The abrupt increase of $T_{2}$ at
$\langle{x}\rangle_{\rm IC} \gtrsim 0.025$
comes from experiment ID 110.

In contrast, the SEA model shows a strong increase in spartyness
as $\langle{x}\rangle_{\rm IC}$ increases for several experiments,
in particular for the HERA combined charm experiment ID 147.
(The CDHSW measurement of $F_2^p$, experiment ID 108,
shows a strong decrease in spartyness;
but we note that that experiment is already anomalously well-fit,
even for no intrinsic charm.)
Again, this is consistent with the rapid rise of
$\chi^{2}$ versus $\langle{x}\rangle_{\rm IC}$
in Figure \ref{fig:chisqVxic} at $\langle{x}\rangle_{\rm IC} \approx 0.015$;
but Figure \ref{fig:spartyall} is more informative,
because it shows which experiments are responsible for the
rapid rise in $\chi^{2}$,
i.e., which experiments conflict with the large IC.
In fact, it is the HERA combined charm experiment 147
that gives the dominant contribution to the Tier-2 penalty
for the SEA model, and determines the limit on
charm momentum fraction from Fig.~\ref{fig:chisq+T2Vxic}.

Overall, we can use the spartyness plots to understand which data sets
can or cannot be fit by a particular model of intrinsic charm.
For instance, the BHPS model actually gives a \emph{better} global fit
for $\langle{x}\rangle_{\rm IC} \sim 0.01$ because
that significantly lowers the spartyness of the BCDMS $F_2^p$ data set 101;
at the same time it does not conflict with HERA inclusive charm data set 147.
Evidently, intrinsic charm at large $x$ at $Q_{c}$
does not conflict with charm production at
HERA (low $x$ and large $Q$),
while it can improve the agreement with the BCDMS data.
On the other hand any increase in the SEA model intrinsic charm
(predominantly low $x$ charm) worsens the fit to both data sets
101 and 147.

\clearpage

\section{Predictions for the LHC
\label{sec:PREDICTIONS}}

\begin{figure}[tbh] \begin{center}
$
\begin{array}{c}
\includegraphics[width=0.39\textwidth]{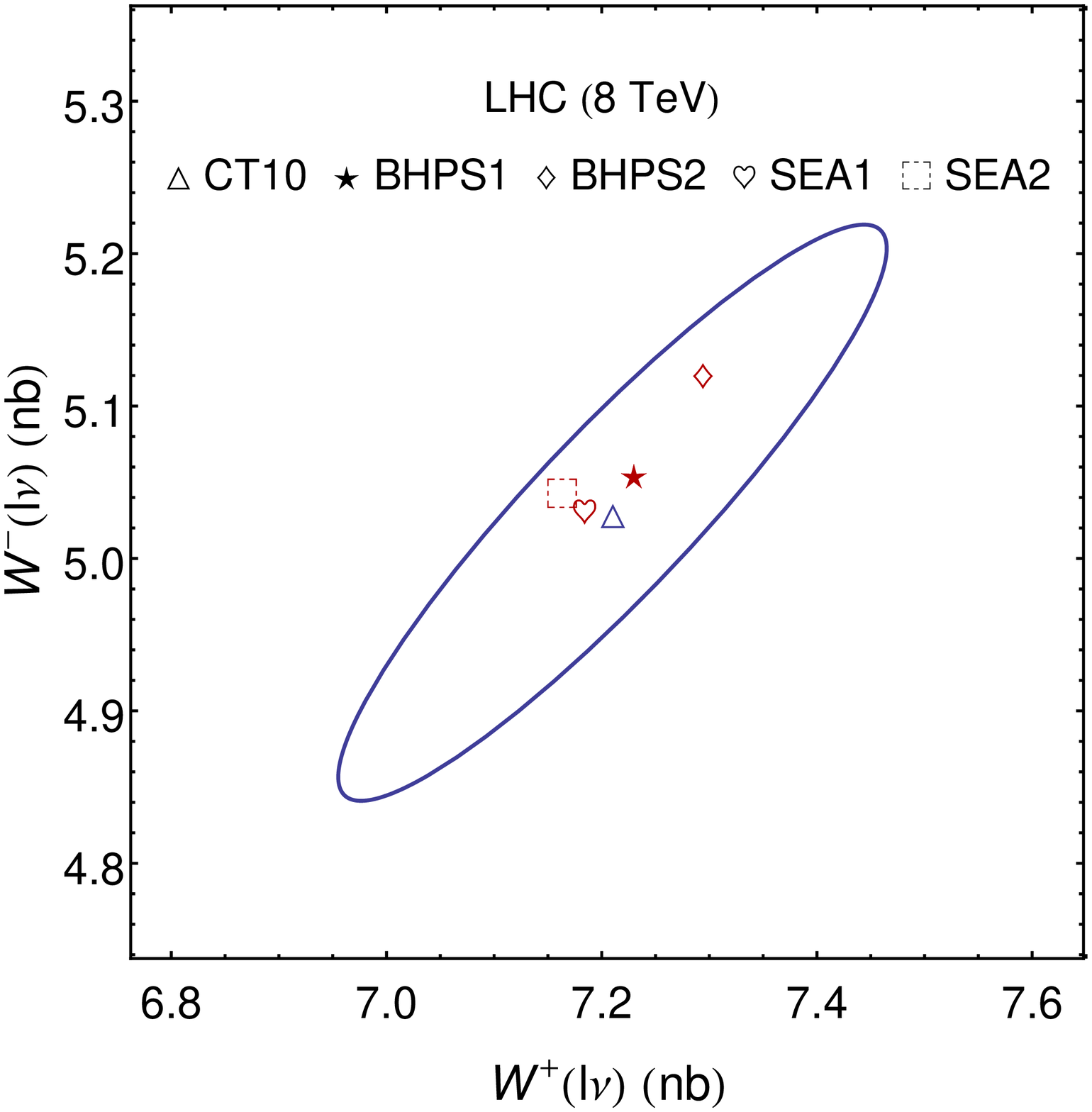}
\hspace{0.3in}
\includegraphics[width=0.4\textwidth]{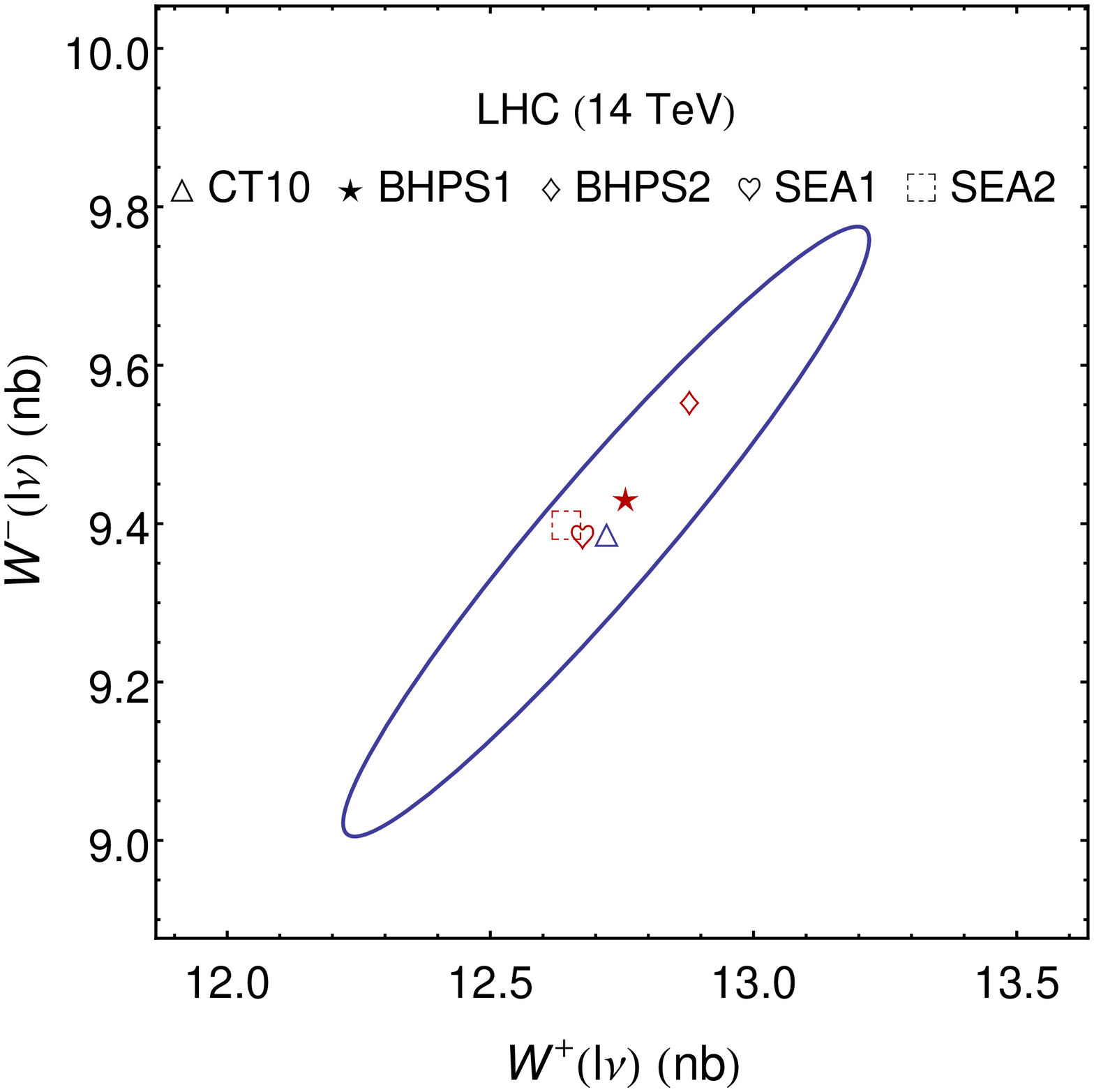}
\end{array}
$
\end{center}
\caption{Correlation plot for the predictions of $W^-$ and $W^+$ boson production
cross sections at the LHC with $\sqrt{S} = $ 8 TeV and 14 TeV.
\label{fig:wpwm}}
\end{figure}

\begin{figure}[tbh] \begin{center}
$
\begin{array}{c}
\includegraphics[width=0.4\textwidth]{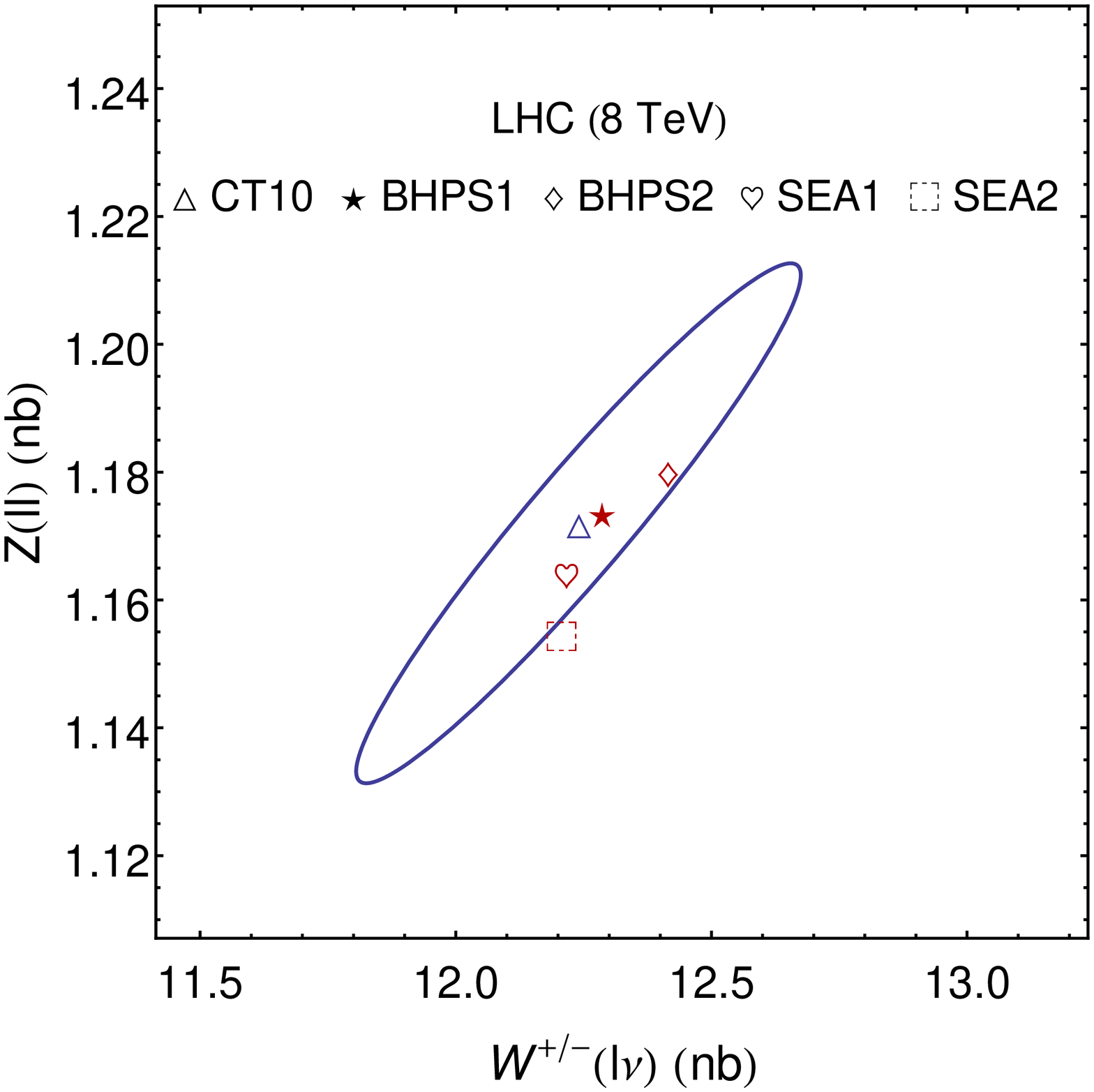}
\hspace{0.3in}
\includegraphics[width=0.4\textwidth]{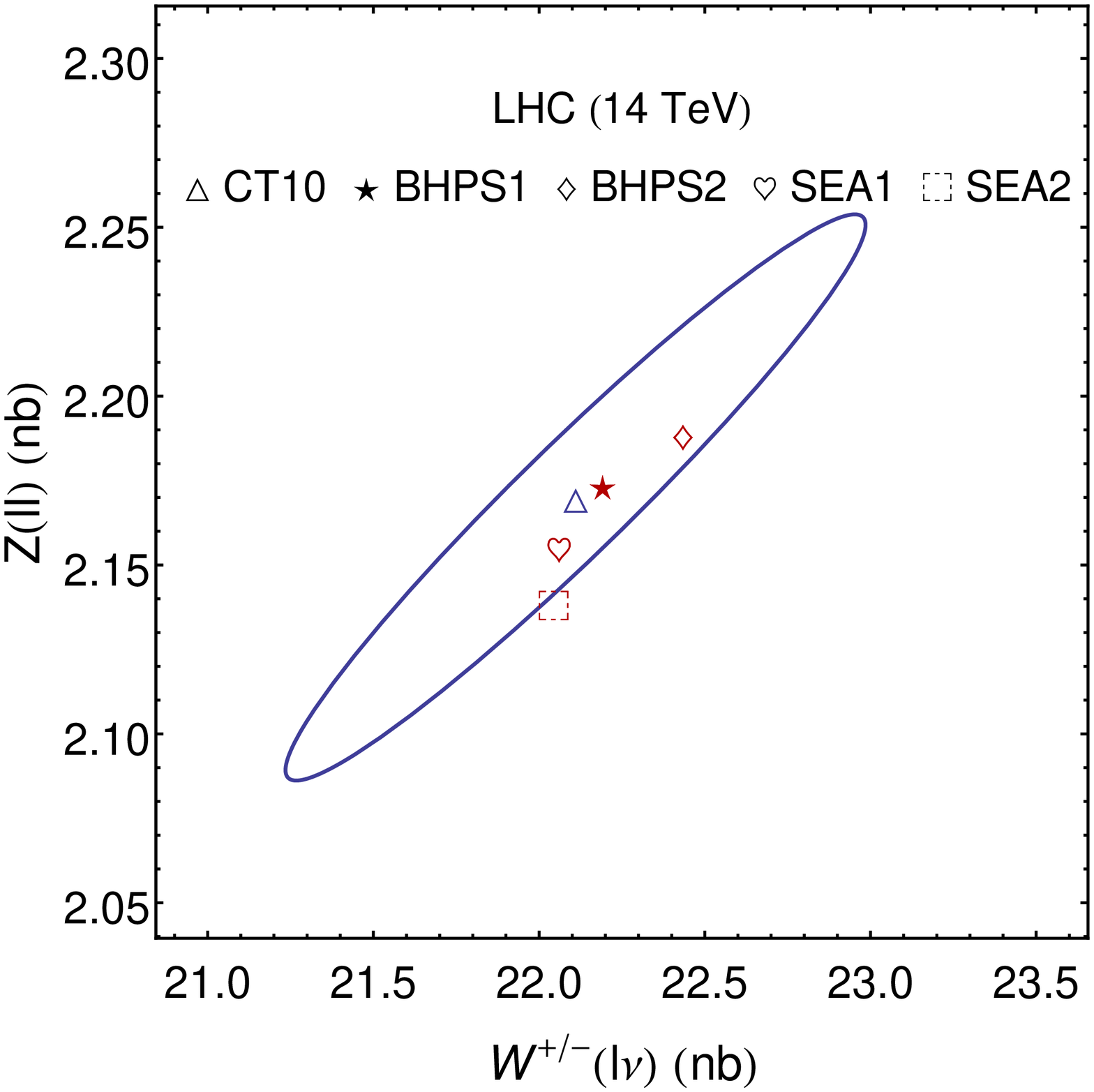}
\end{array}
$
\caption{Correlation plot for the predictions of $Z$ and $W^{\pm}$ boson production
cross sections at the LHC with $\sqrt{S} = $ 8 TeV and 14 TeV.
\label{fig:zgws}}
\end{center} \end{figure}

\begin{figure}[tbh] \begin{center}
$
\begin{array}{c}
\includegraphics[width=0.4\textwidth]{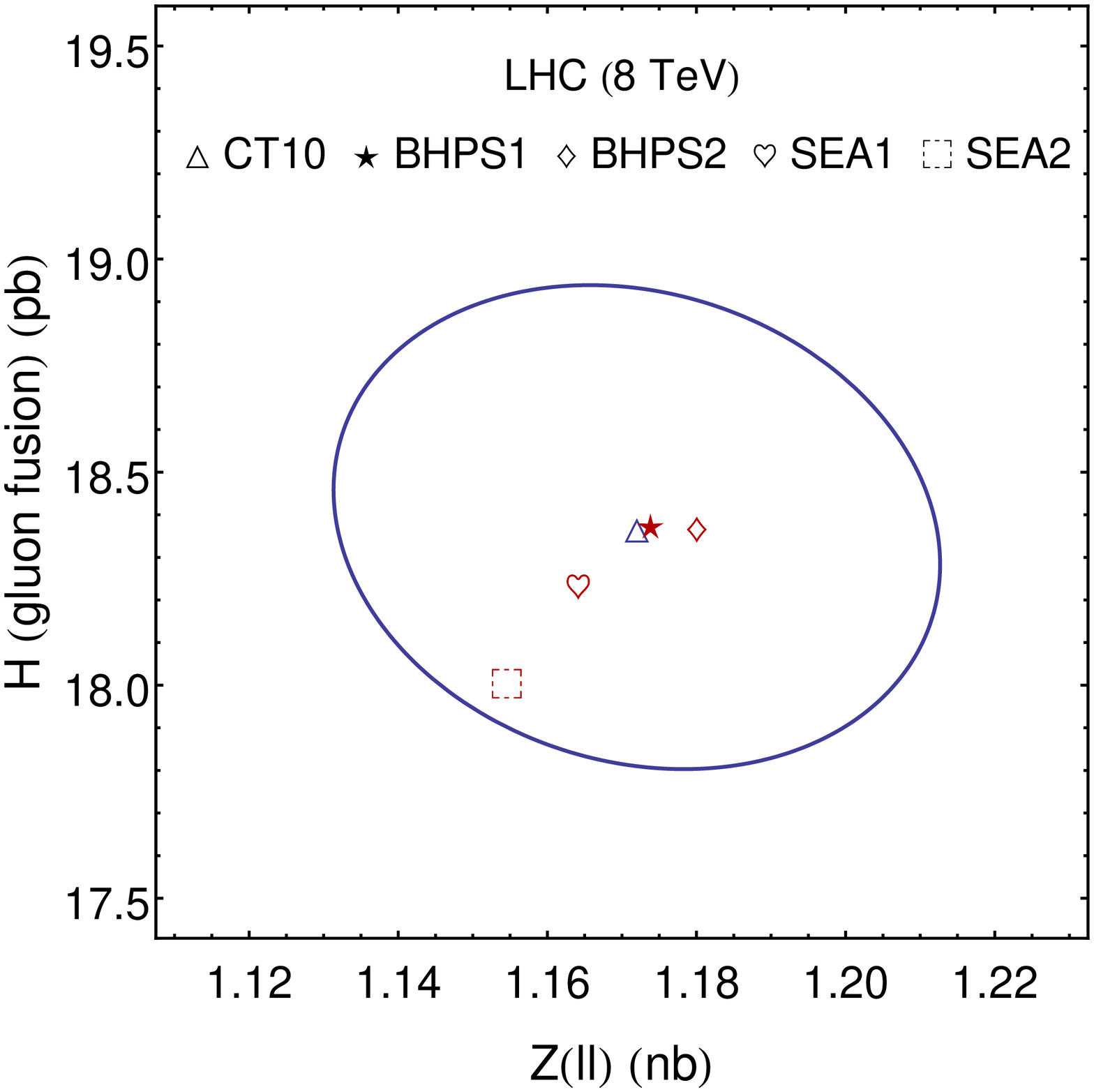}
\hspace{0.3in}
\includegraphics[width=0.39\textwidth]{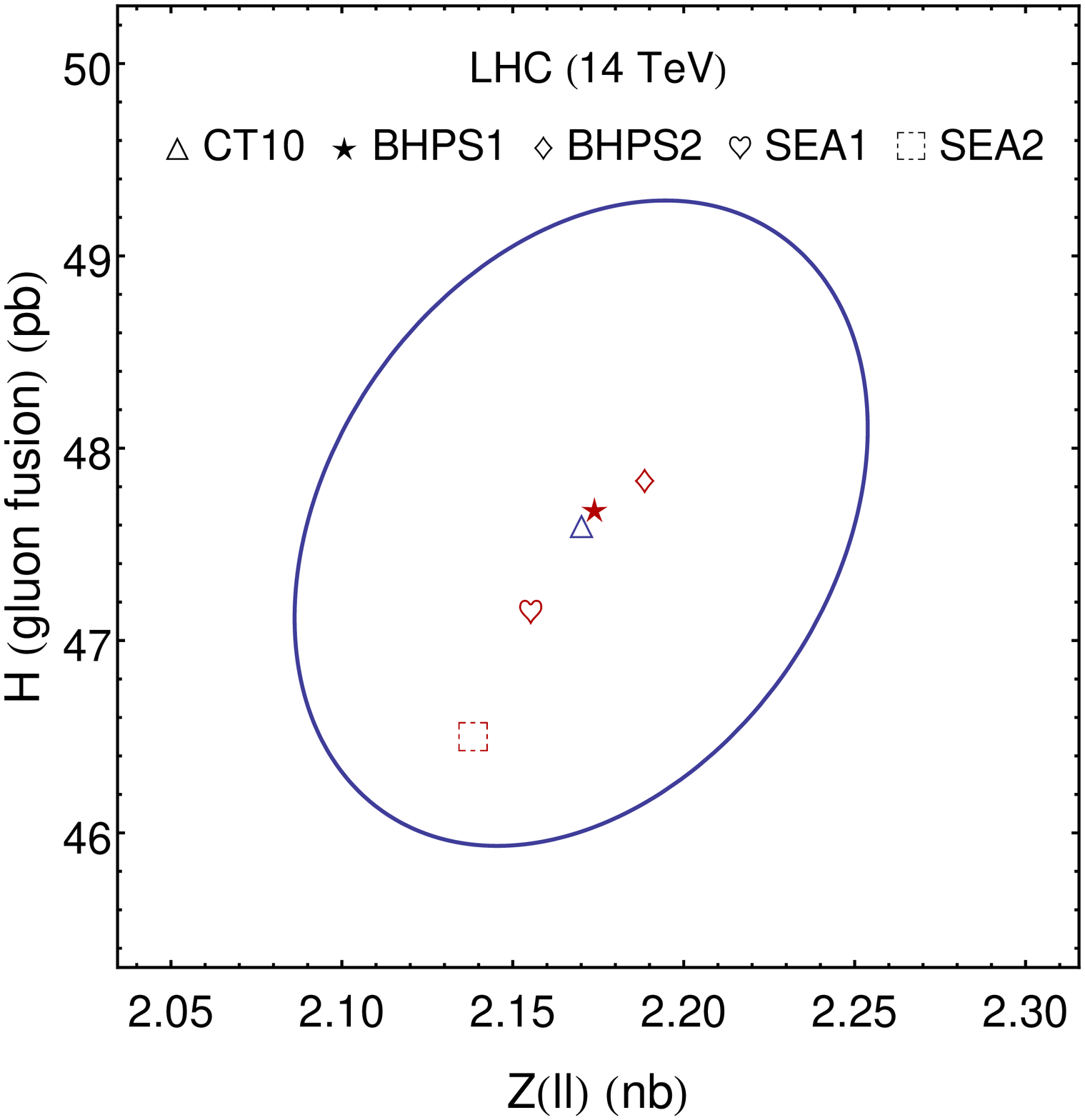}
\end{array}
$
\caption{Correlation plot for the predictions of Higgs and $Z$ boson production
cross sections at the LHC with $\sqrt{S} = $ 8 TeV and 14 TeV.
\label{fig:zgh}}
\end{center} \end{figure}

\begin{figure}[tbh] \begin{center}
$
\begin{array}{c}
\includegraphics[width=0.39\textwidth]{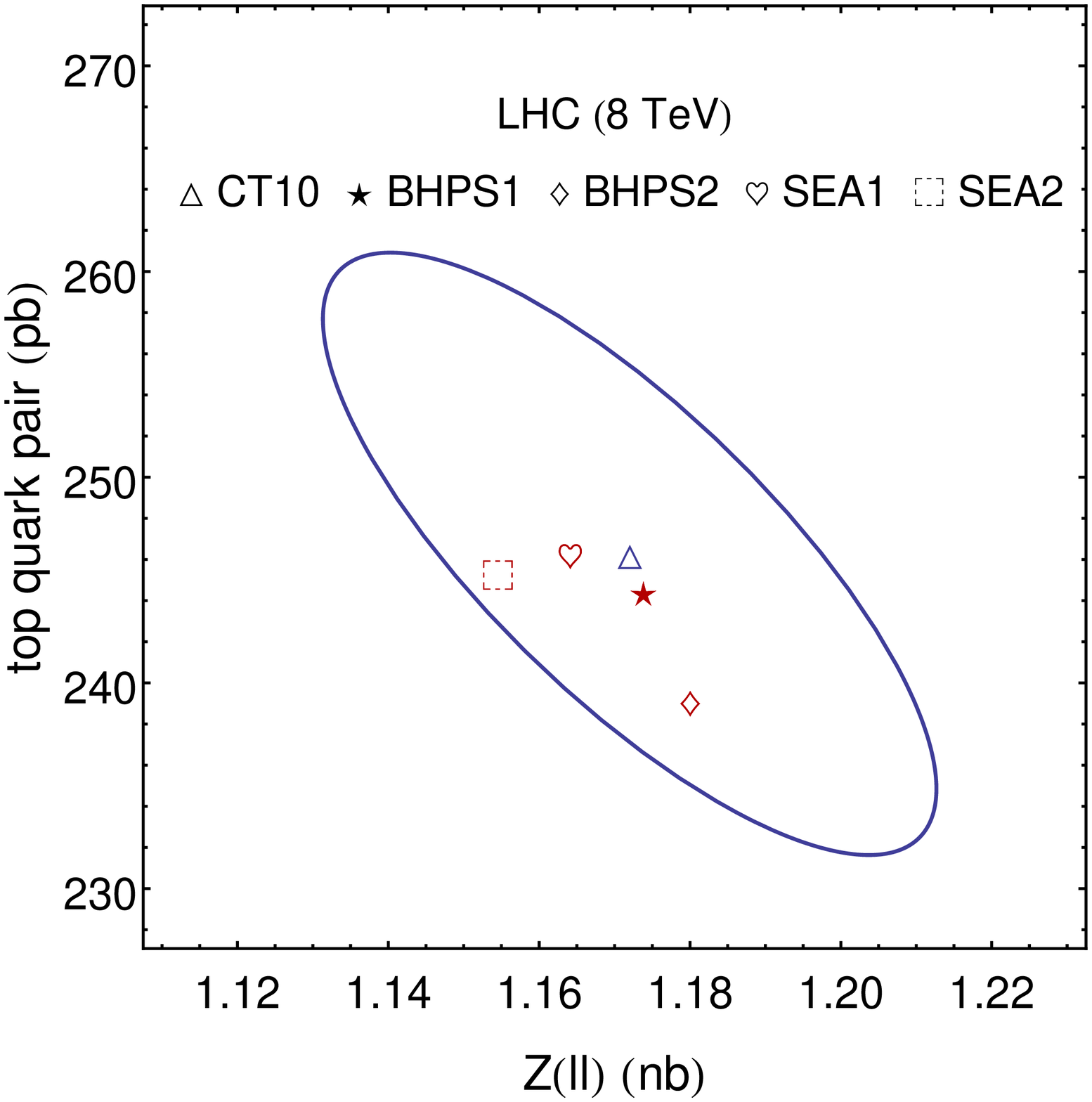}
\hspace{0.3in}
\includegraphics[width=0.4\textwidth]{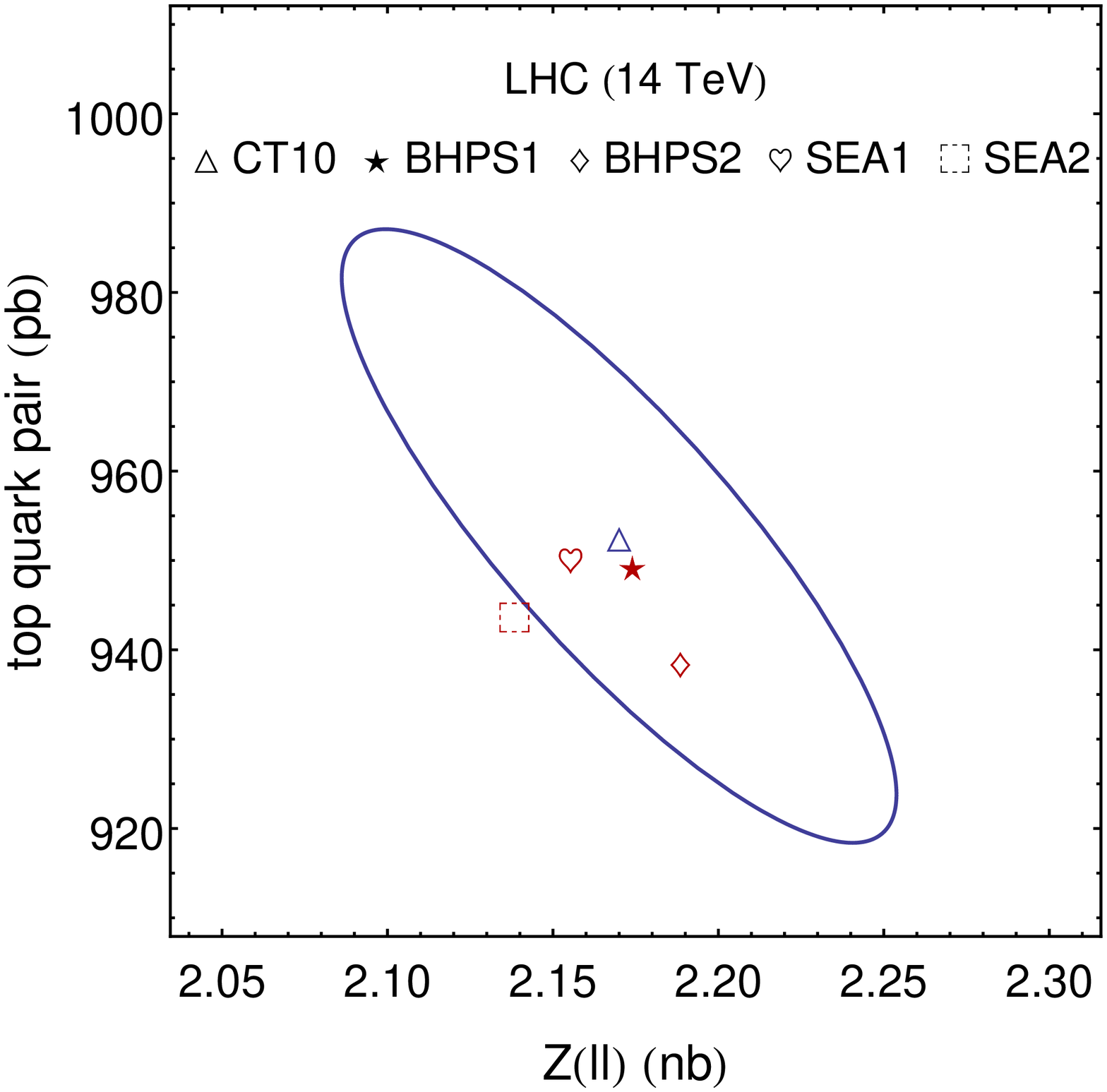}
\end{array}
$
\caption{Correlation plot for the predictions of top quark pair and $Z$ boson production
cross sections at the LHC with $\sqrt{S} = $ 8 TeV and 14 TeV.
\label{fig:zgtt}}
\end{center} \end{figure}

The inclusion of intrinsic charm changes the charm quark
distributions at high $Q$ values significantly, as well as affects the
gluon and other quark PDFs through the PDF correlations in the global fit as
shown in Section~\ref{sec:RESULTS}. Thus it may have impact on
collider observables~\cite{ct66}. For example, in
Figs.~\ref{fig:wpwm}-\ref{fig:zgtt} we show the predictions of the
NNLO total cross sections for $W$ and $Z$ boson production, Higgs boson
production through gluon fusion, and top quark pair production at the LHC at
$\sqrt{S}=8$ and 14 TeV. The NNLO cross sections for $W$ and $Z$
boson production are computed with
FEWZ2.1~\cite{fewz1,fewz2}. The NNLO cross sections
for Higgs boson and top quark pair production are obtained from
iHix1.3~\cite{ihixs} and
Top++2.0~\cite{toppp1,toppp2}, with $m_h=125\
{\rm GeV}$, $m_t=173.3\ {\rm GeV}$ and the QCD scales set to the
corresponding mass values. For each pair of total cross sections, we
show the central predictions from CT10 with the 90\% C.L. PDF tolerance
ellipse as well as predictions from the four examples with intrinsic charm.

In Figs.~\ref{fig:wpwm}-\ref{fig:zgtt}, generally the predictions
from the chosen IC models differ from the
central predictions of CT10 by less than 2\%, which is smaller
than the PDF uncertainties of CT10 but may not be negligible for
precise predictions for the LHC. For the SEA models, the cross sections
at the LHC are almost uniformly smaller,
due to the reduction in momentum fraction
remaining for the other non-charm partons.
The only exception is $W^-$ production
through the process $\bar{c}s\rightarrow W^-$, which benefits from the
increased $\bar{c}$ distribution.
The production of Higgs bosons and top quark pairs,
in particular, are reduced for nonzero IC SEA, due to the reduction in gluons
and light sea quarks in the relevant regions of $x$ and $Q$, as seen previously
in Figs.~\ref{fig:gluratios} and \ref{fig:ubrratios}.

For the BHPS models, the situation is more complicated, since the gluon and
light sea quark PDFs are increased in the region $x\lesssim0.1$  in the BHPS models.
In this case the production of $Z$ and $W^\pm$ (independent of sign)
are increased and the production of top quark pairs is decreased
by an increase of IC in the BHPS models.
The production of Higgs bosons is
fairly insensitive to the amount of IC for the BHPS model.
Also it is interesting to see
from Figs.~\ref{fig:wpwm}, \ref{fig:zgws} and \ref{fig:zgtt} that
the predictions from the BHPS models follow similar (anti) correlations
as CT10 since they distribute along the diagonal direction
of the ellipse,  in distinction to the SEA models.

\begin{figure}[tbh] \begin{center}
\includegraphics[width=0.47\textwidth]{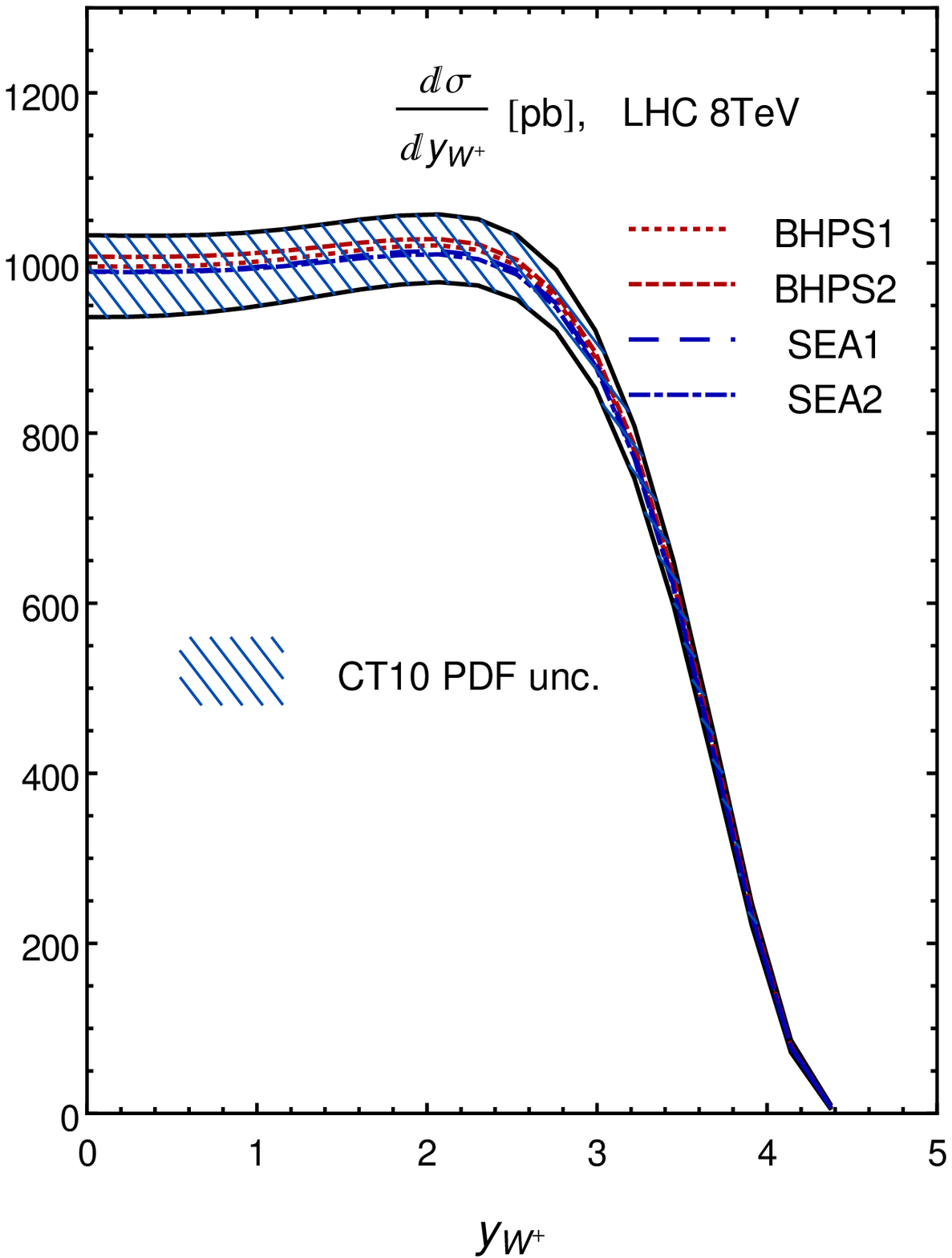}
\includegraphics[width=0.47\textwidth]{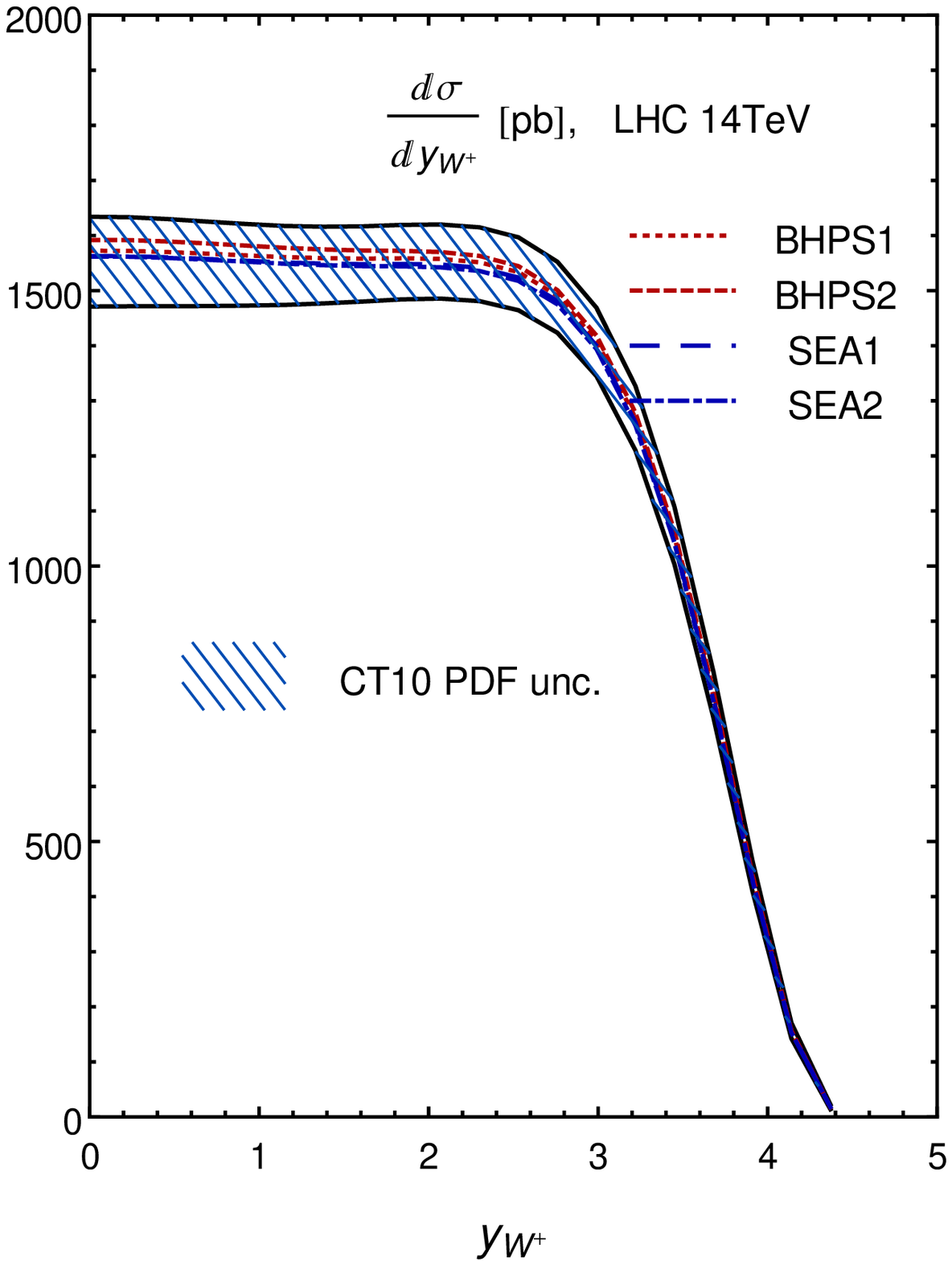}
\caption{Rapidity distribution of the $W^+$ boson at the LHC with $\sqrt{S} = $ 8 TeV and 14 TeV.
\label{fig:rapwp}}
\end{center} \end{figure}

\begin{figure}[tbh] \begin{center}
\includegraphics[width=0.47\textwidth]{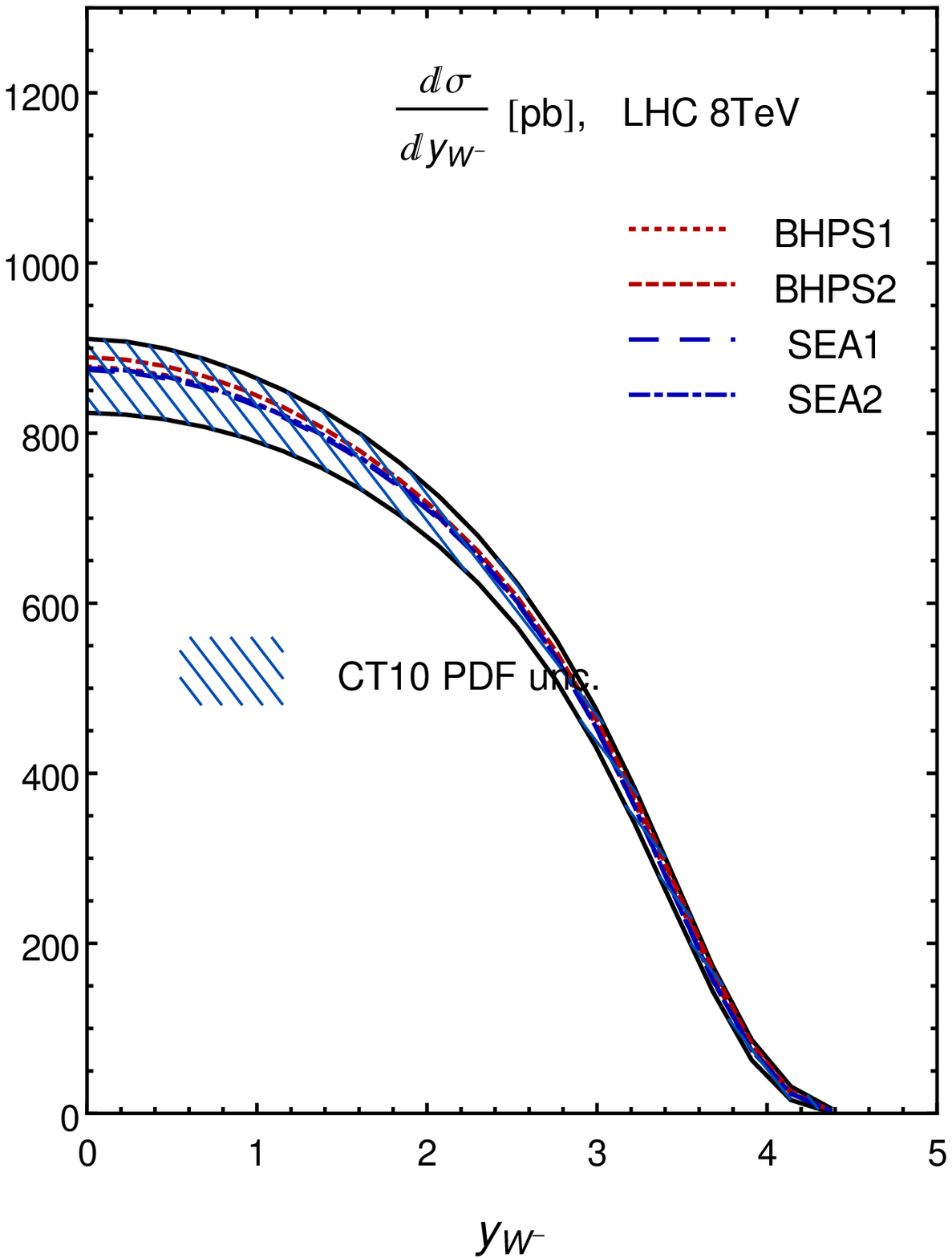}
\includegraphics[width=0.47\textwidth]{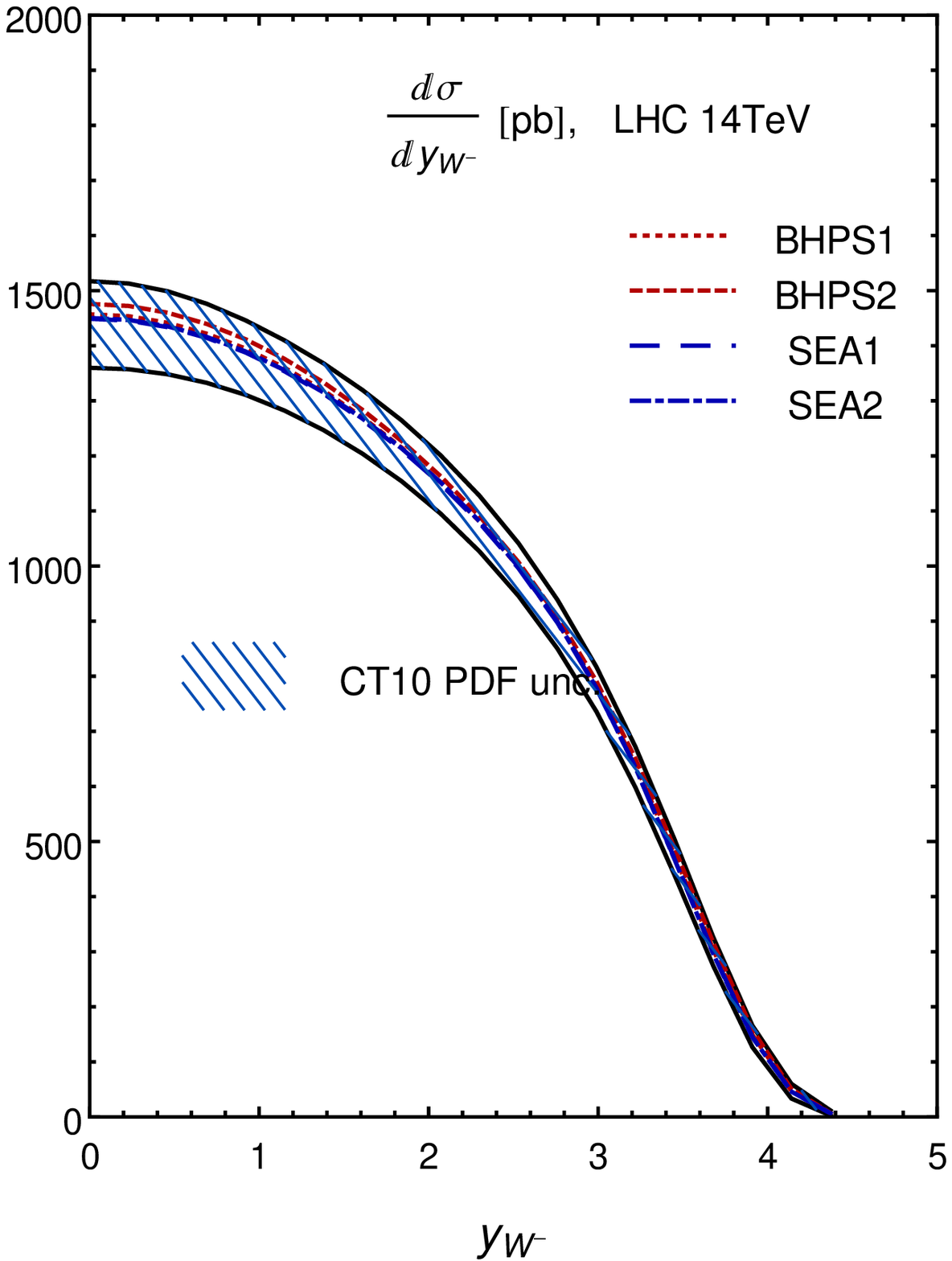}
\caption{Rapidity distribution of the $W^-$ boson at the LHC with $\sqrt{S} = $ 8 TeV and 14 TeV.
\label{fig:rapwm}}
\end{center} \end{figure}

\begin{figure}[tbh] \begin{center}
\includegraphics[width=0.47\textwidth]{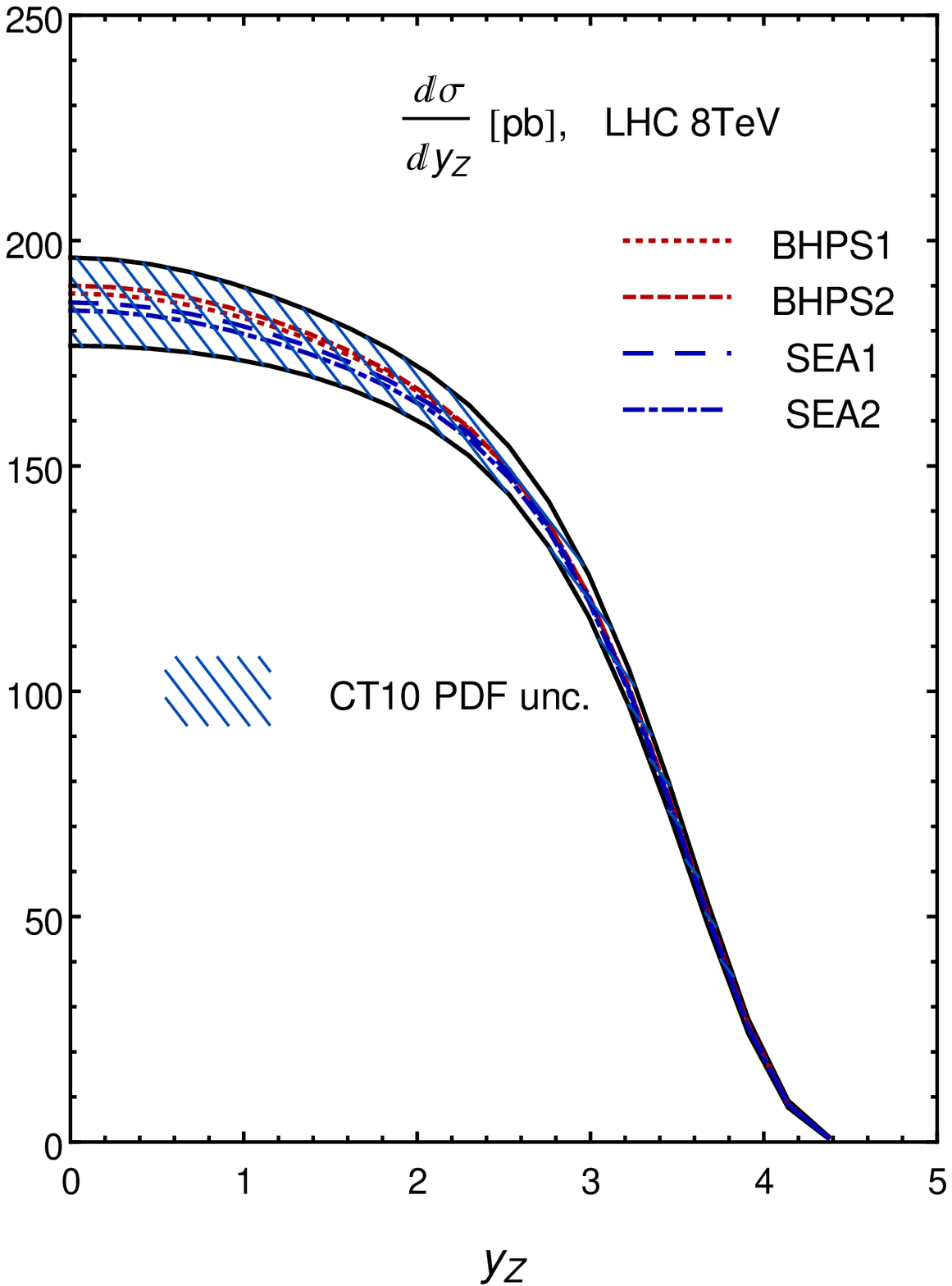}
\includegraphics[width=0.47\textwidth]{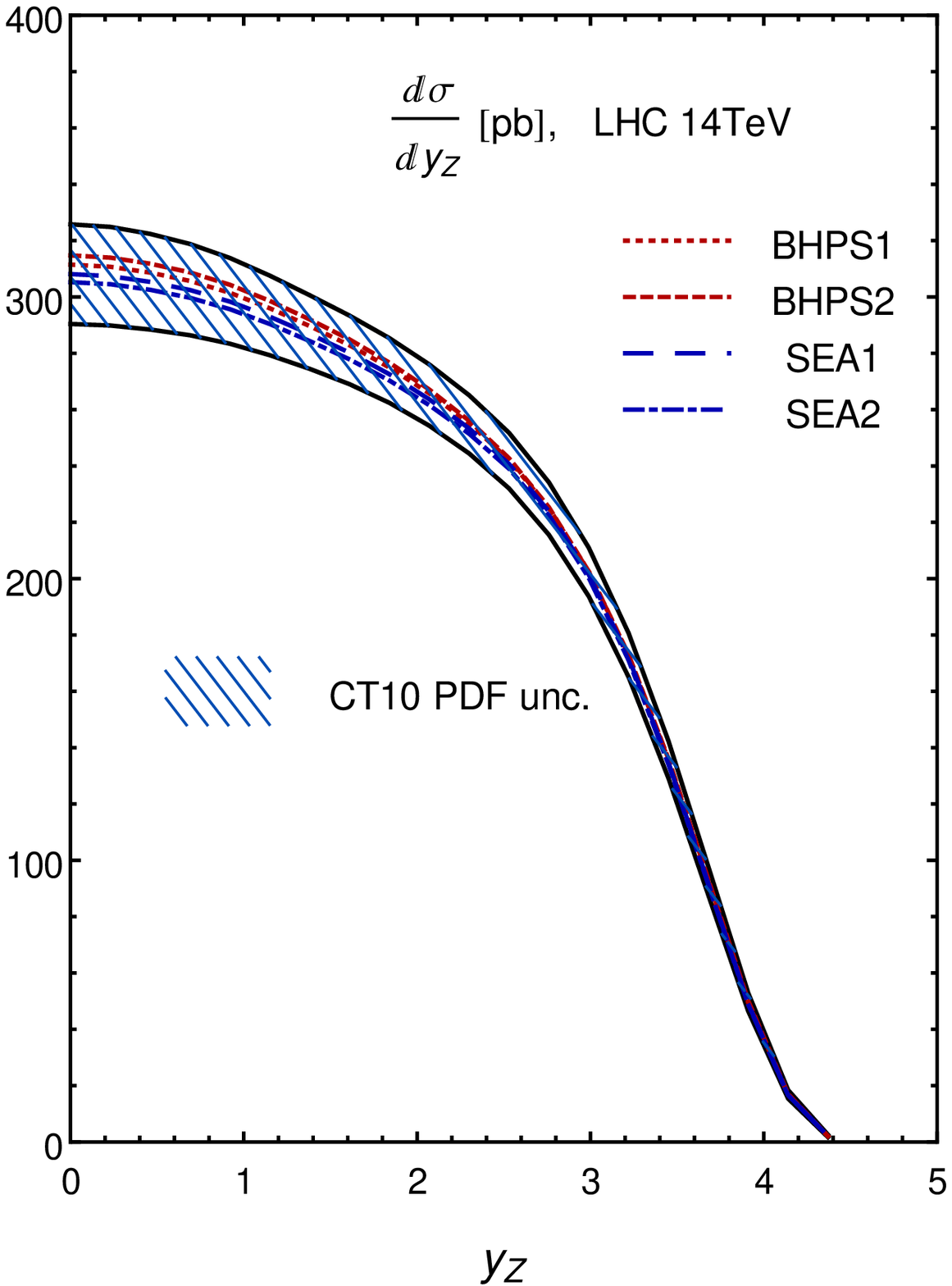}
\caption{Rapidity distribution of the $Z$ boson at the LHC with $\sqrt{S} = $ 8 TeV and 14 TeV.
\label{fig:rapzg}}
\end{center} \end{figure}

\begin{figure}[tbh] \begin{center}
\includegraphics[width=0.47\textwidth]{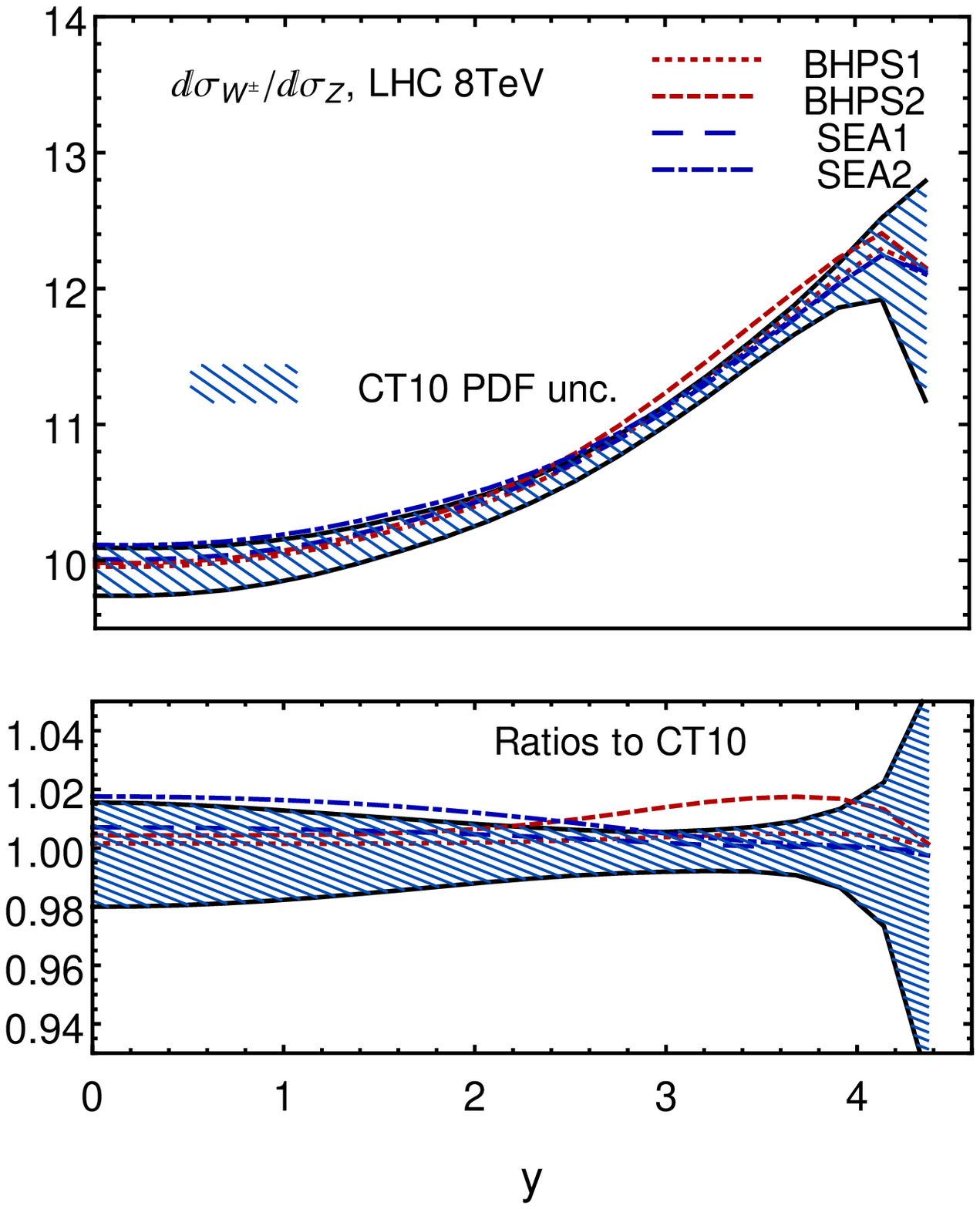}
\includegraphics[width=0.47\textwidth]{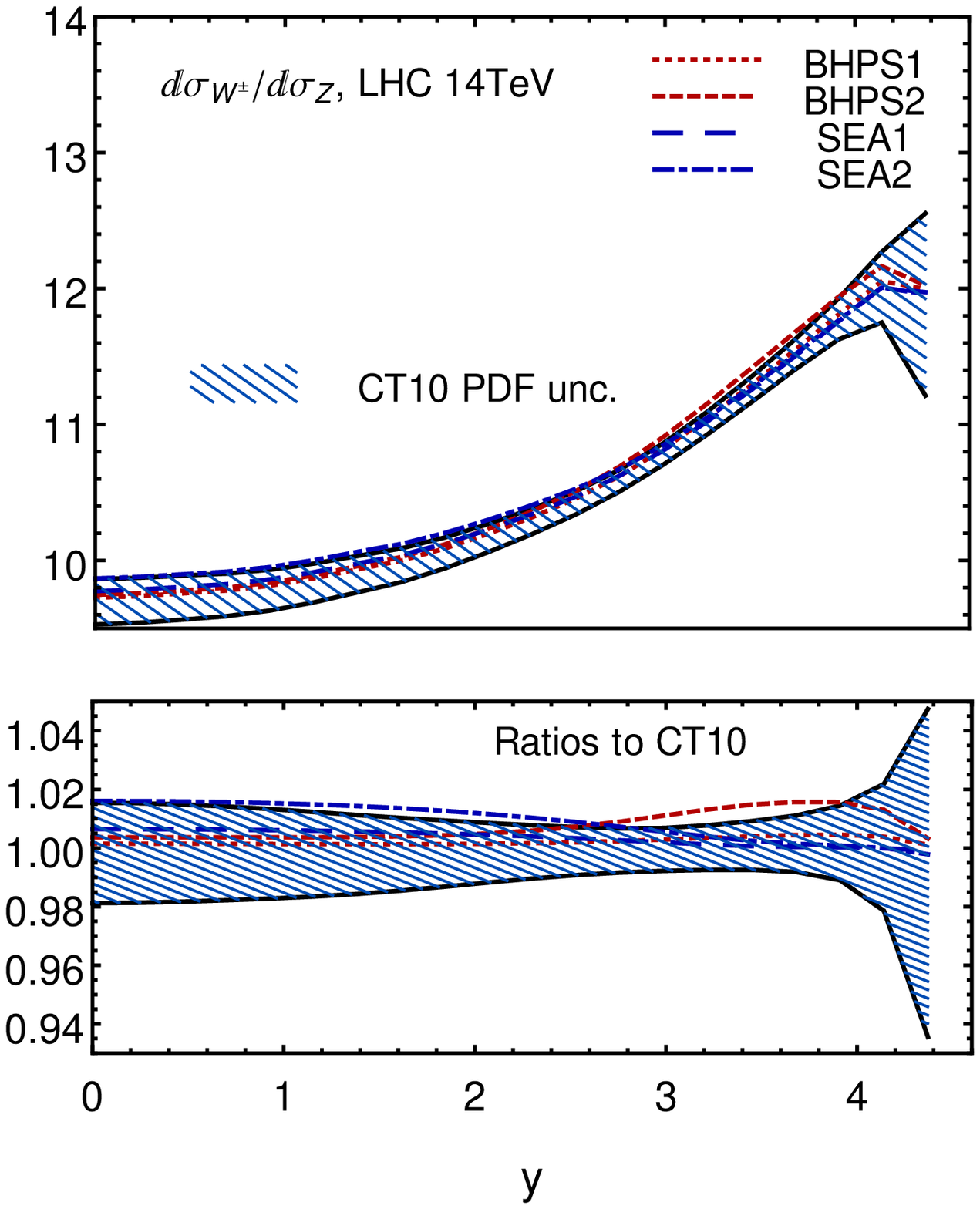}
\caption{The ratio $d\sigma_{W^+ + W^-}\over d\sigma_Z$
at the LHC with $\sqrt{S} = $ 8 TeV and 14 TeV for four IC models at NNLO.
Upper panel: differential distributions;
lower panel: ratios normalized to the CT10 central prediction.
\label{fig:rapra}}
\end{center} \end{figure}

We can also check the effects of the IC on the rapidity distribution
of the vector boson production. Charm quark contributions there have different
shapes compared to the light quarks depending on the momentum profile of the
charm quark of the IC models. Figs.~\ref{fig:rapwp}-\ref{fig:rapzg} give the NNLO predictions
of the rapidity distributions of $W^\pm$ and Z boson productions from CT10 and the
four IC models. They are calculated with the program Vrap0.9~\cite{vrap},
and both the renormalization and factorization scales are chosen to be
the mass of the vector boson.
 The differences
are small among four IC models, and their predictions are all within the CT10 PDF
uncertainties. In Fig.~\ref{fig:rapra} we further plot the ratios of the rapidity distribution,
$d\sigma_{W^{\pm}}/d\sigma_Z$, which presumably are more sensitive to the charm quark
contributions due to the cancellations of the uncertainties from light quark contributions.
This is evident from comparing the size of error band induced
by the PDF uncertainty in Figs.~\ref{fig:rapwp}-\ref{fig:rapzg} to that in Fig.~\ref{fig:rapra}.
We can see that in the small or intermediate rapidity region the predictions of SEA2
lie outdide the CT10 PDF uncertainties, similar to that
 for BHPS2 in the large rapidity region.

\begin{figure}[tbh] \begin{center}
\includegraphics[width=0.47\textwidth]{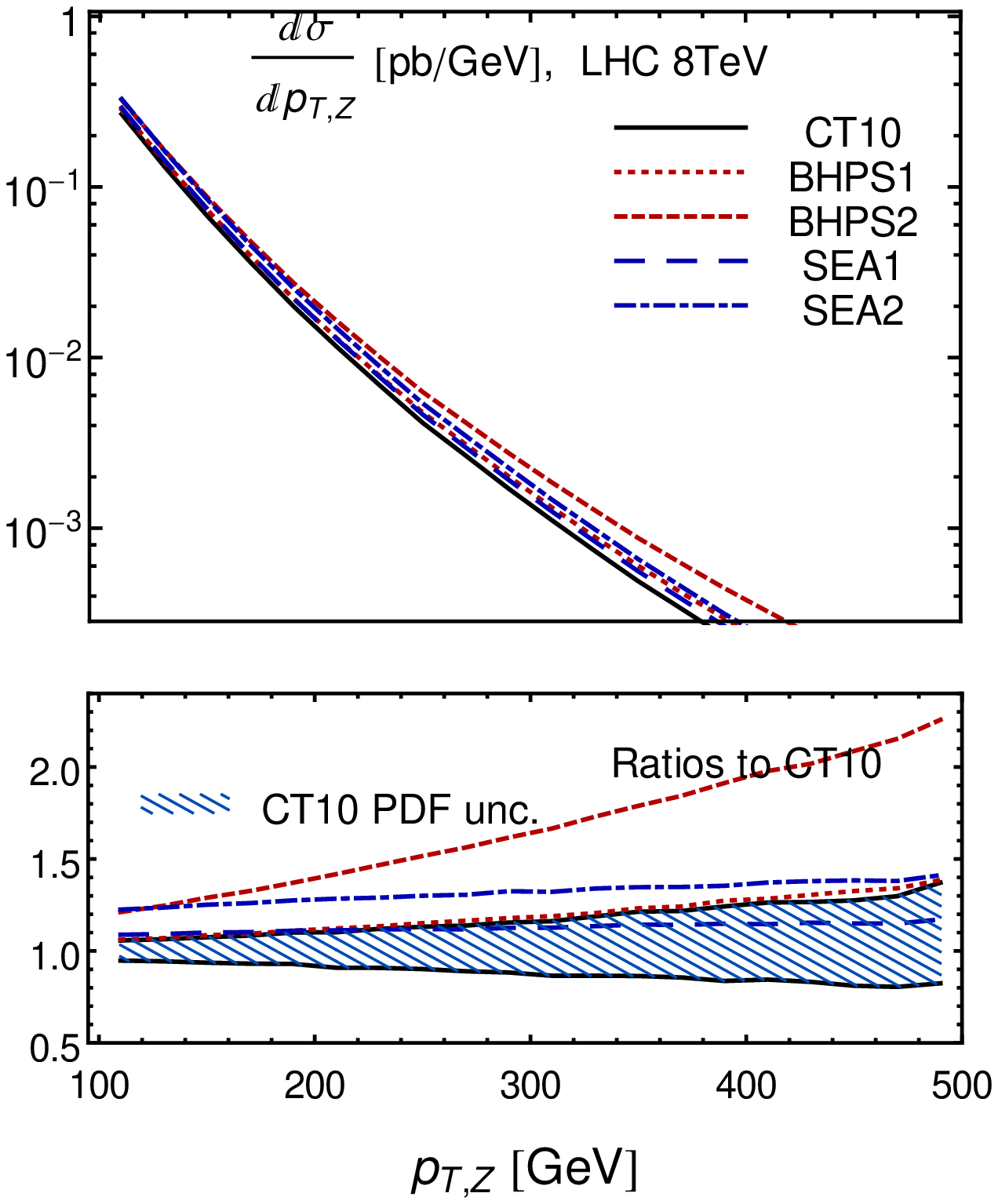}
\includegraphics[width=0.47\textwidth]{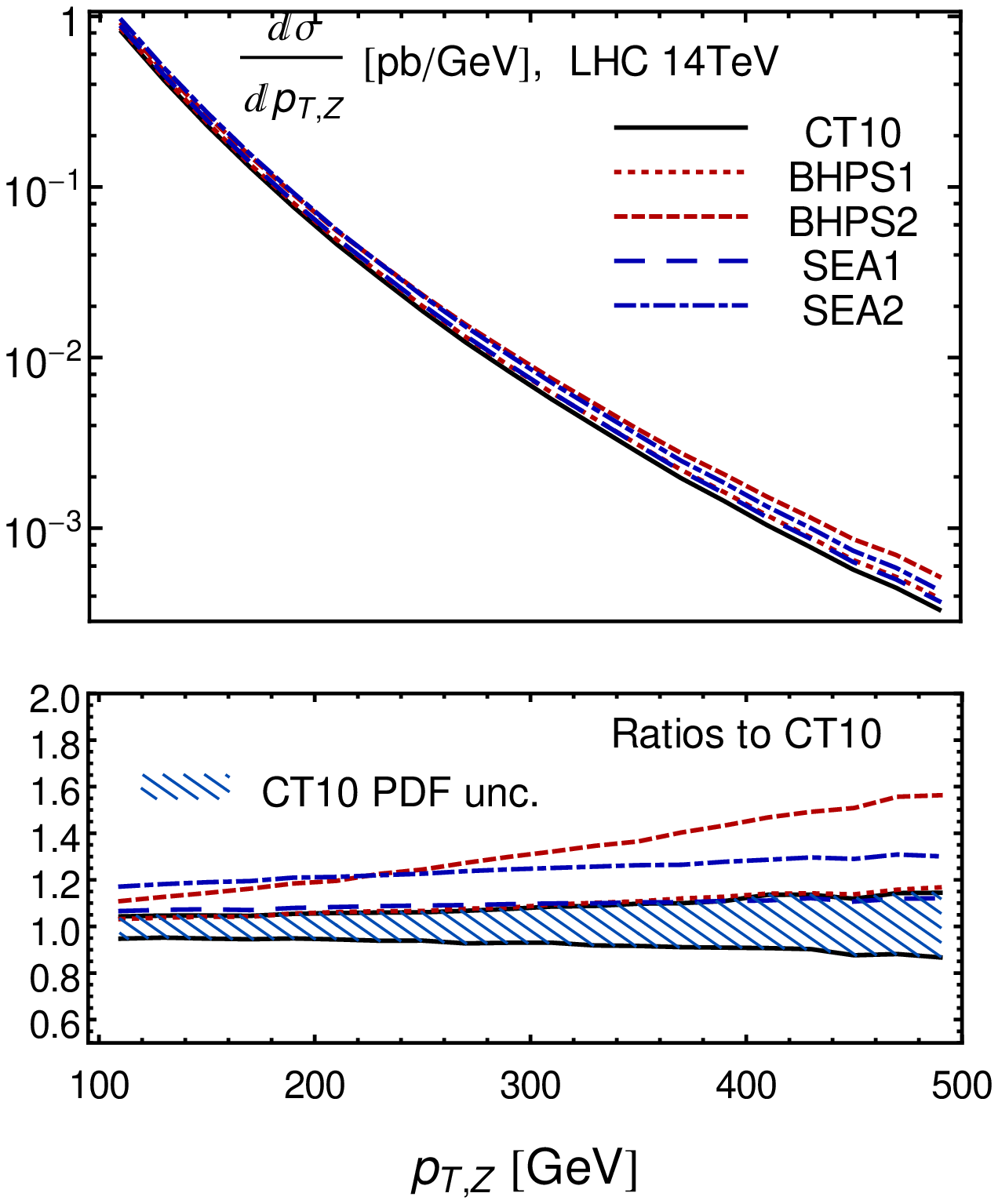}
\caption{Transverse momentum distribution of the $Z$ boson in the production
of $pp\rightarrow Zc$ at the LHC with $\sqrt{S} = $ 8 TeV and 14 TeV.
Upper panel: differential cross sections;
lower panel: ratios normalized to the CT10 central prediction.
\label{fig:zqlhc}}
\end{center} \end{figure}

As already pointed out in our previous IC study~\cite{plt},
and recently in Refs.~\cite{zq1,zq2},
the partonic process $g\,+\,c\,\rightarrow\,\gamma/Z\,+\,c$
is directly sensitive to the initial state charm distribution,
although precision measurements could be experimentally
challenging at the LHC.
Figure \ref{fig:zqlhc} shows predictions of the differential
cross sections of an on-shell $Z$ boson production in association with a
charm quark at the LHC from CT10 and the IC models.
The matrix element calculations of the process are only available up to
NLO in QCD~\cite{zq3} and are implemented in program MCFM~\cite{mcfm}.
We simply convolute them with our NNLO PDFs in order to show the relative
changes of the cross sections in the presence of IC.
Figure \ref{fig:zqlhc} gives the transverse momentum distribution
of the $Z$ boson with kinematic cuts of $p_{T} > 50\,{\rm GeV}$,
and $|\eta|<2.1$ applied on the charm quark jet using the anti-$k_T$
algorithm with the radius parameter $R=0.5$.
In this calculation, both the renormalization and factorization
scales are chosen to be the scale sum of the transverse momenta
of final state particles. We have checked that in the
large $p_T$ region, the theoretical uncertainty induced by
varying these scales simultaneously by a factor of 2 is much
smaller than the error induced by the CT10 error PDFs.

We can clearly see that in the large $p_{T,Z}$ region,
predictions from the fits with IC models deviate sigificantly
from CT10 as a direct result of the charm distribution changes
shown in Fig.~\ref{fig:zqlhc}.
Thus, it shows a potential for discriminating the possible IC models.
In particular, at large values of $p_{T,Z}$ it is very sensitive to the presence
of IC in the BHPS models, since they have a large enhancement at large $x$.
However, the cross sections are small, and there are further suppressions
from the decay branching ratios of the $Z$ boson, as well as the charm quark
jet-tagging efficiency.
A detailed study of the feasibility of this process at the
LHC is needed, but it is beyond the scope of the current work.

\clearpage

\section{Correlations between the charm mass and intrinsic charm
\label{sec:McDependence}}

The curves in Fig.~\ref{fig:cpdf} show that the charm distribution in the SEA model
is very similar in shape to the charm distribution of CT10 with no IC.
It mostly differs in overall normalization.
This then begs the question whether it can be distinguished
from some other physics that may produce more $c\bar{c}$ radiation,
such as a change in the charm quark mass, $m_c$.
In this section we investigate the correlation between $m_c$
and IC in the SEA model.

A lighter charm quark mass could mimic the effects of intrinsic charm,
because with a smaller mass,
the charm PDF turns on sooner, resulting in more charm at a given value of $Q$. Conversely, a larger
value of the charm mass would result in a smaller charm content at a given value of $Q$, which could
then be made up with some intrinsic charm.
This possibility is intriguing because the standard value of the
charm mass $m_c^{\rm pole}=1.3$ GeV,
which was used in the CT10 analysis~\cite{ct10nn} and was
shown to be favored by the global analysis data~\cite{smucharm},
is smaller than the world average given
by the Particle Data Group (PDG)~\cite{pdg},
which was obtained using a mostly orthogonal set of data.
The PDG gives a world average value of $m_c(m_c)=1.275$ GeV,
which translates into pole mass values of
$m_c^{\rm pole}=1.46$ GeV and $m_c^{\rm pole}=1.67$ GeV
at one-loop and two-loop conversions, respectively.
Thus, one may wonder if the lower value of $m_c$ used in the global analysis
is actually hiding some evidence for intrinsic charm.

\begin{figure}[H]
\begin{center}
\includegraphics[width=0.8\textwidth]{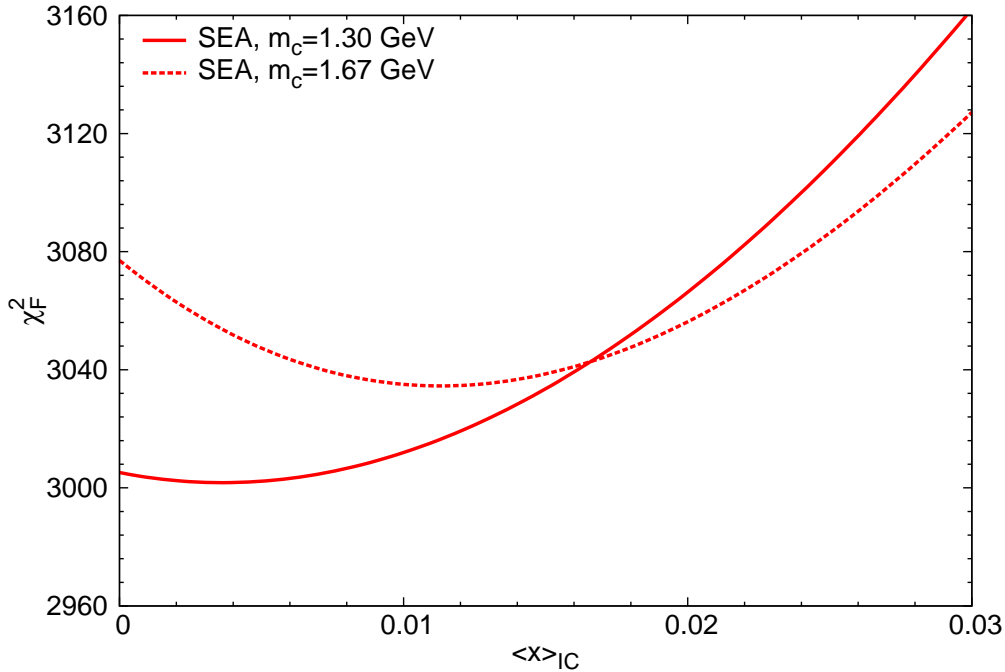}
\end{center}
\caption{The global chi-square function
$\chi_{F}^{2}$ versus charm momentum fraction, $\langle{x}\rangle_{\rm IC}$,
for the SEA model with $m_c=1.3$ GeV and $m_c=1.67$ GeV.
\label{fig:chisqVxic_mc}}
\end{figure}

Rather than performing a combined fit of $m_c$ and IC,
we shall just demonstrate the effect by re-running the
analysis of section~\ref{sec:RESULTS},
using the larger value of $m_c=1.67$ GeV.
We emphasize that we are not advocating this large value of the charm mass,
but are just using it to observe the correlation between fits of the charm mass and intrinsic charm.
In Fig.~\ref{fig:chisqVxic_mc} we plot the results of the
global fitting for $\chi^2_F$ versus the intrinsic charm content,
 $\langle{x}\rangle_{\rm IC}$, for the SEA model,
 both for our standard charm mass $m_c=1.3$ GeV
and for the larger value of $m_c=1.67$ GeV.
We keep the same initial scale for the light partons, $Q_0=1.295$ GeV,
for both choices of $m_c$.
As seen previously, the dependence of $\chi^2_F$ on
$\langle{x}\rangle_{\rm IC}$ for $m_c=1.3$ GeV is quite flat
for small charm content and then begins to rise.
The curve looks like a quadratic function with a minimum
close to $\langle{x}\rangle_{\rm IC}=0$.
This is consistent with the fact the $m_c=1.3$ GeV is near the best fit
of the charm mass with zero IC to the global analysis data.
On the other hand, for $m_c=1.67$ GeV,
the $\chi^2_F$ begins at a higher value for zero IC
and then noticeably decreases to a minimum at around
$\langle{x}\rangle_{\rm IC}=0.01$.
Thus, if a larger value of the charm mass could be decisively
shown to be required,
then the global analysis would prefer nonzero IC,
even for the SEA model, although this preference is not currently
statistically significant.

\begin{figure}[tbh]
\begin{center}
$
\begin{array}{c}
\includegraphics[width=0.8\textwidth]{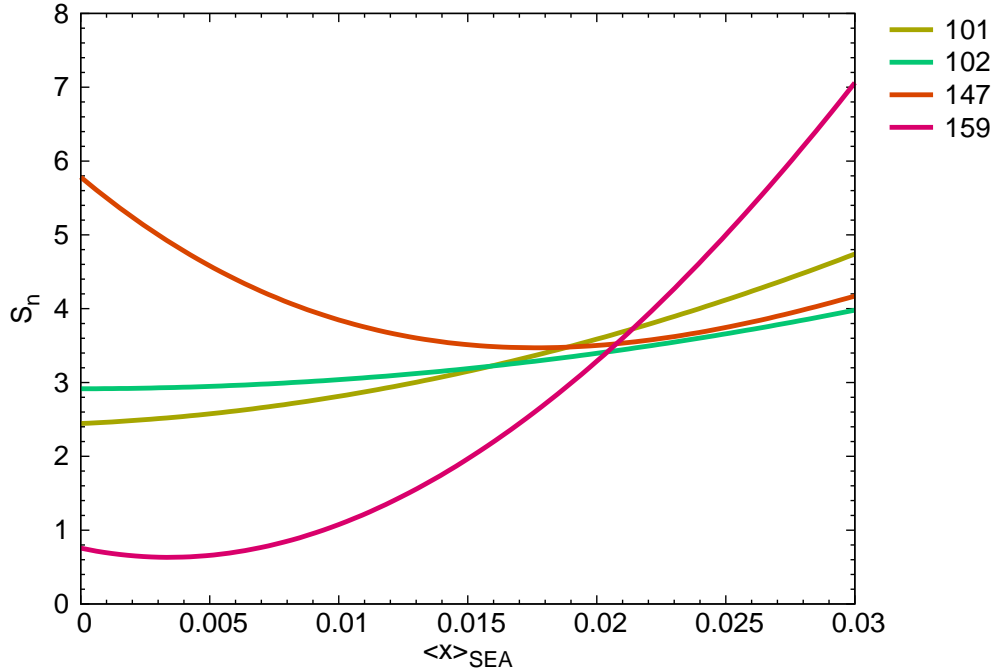}\\
\end{array}
$
\caption{Comparing the ``spartyness'' $S_{n}$ as a function of
$\langle{x}\rangle_{\rm IC}$,
showing the data sets for which $S_{n}$ changes the most significantly
for the SEA model with $m_c=1.67$ GeV.
\label{fig:spartysig_mc}}
\end{center} \end{figure}

We note, however, that the overall value of $\chi^2_F$ is still worse for $m_c=1.67$ GeV and $\langle{x}\rangle_{\rm IC}=0.01$ than for $m_c=1.3$ GeV with zero intrinsic charm.
We investigate this further in Fig.~\ref{fig:spartysig_mc},
where we compare the ``spartyness'' $S_n$ as a function of $\langle{x}\rangle_{\rm IC}$ for the same set of experiments
that were sensitive to the SEA model in Fig.~\ref{fig:spartysig} for $m_c=1.3$ GeV. The main thing to note here is that the spartyness for data set 147,
combined charm production at HERA,
is significantly worse for $m_c=1.67$ GeV than for $m_c=1.3$ GeV.
For $m_c=1.67$ GeV and zero IC,
the spartyness of data set 147 is more than 5,
which indicates a very poor fit to the data.
Increasing the IC content makes the fit better for this charm mass,
but the spartyness is still a good bit higher than that
with $m_c=1.3$ GeV and no IC.
In addition, the fit to the combined HERA1 DIS data set 159 worsens
as the charm momentum fraction increases for the SEA model with $m_c=1.67$ GeV.

\section{Conclusions
\label{sec:CONCLUSIONS}}

The hypothesis of intrinsic charm (IC) is a long-standing question in
high-energy physics, combining QCD theory (both perturbative and
nonperturbative) and phenomenology.\cite{bhps}
The purpose of this paper is to reassess the theory based on
up-to-date global analysis of QCD at the NNLO approximation.

The final conclusion of this work is that the wide range of short-distance
processes that are commonly used in global analysis can be described accurately
without an IC component of the proton;
however, they do not rule out a small IC component.
Quantitatively, we can construct PDFs that are acceptable
fits to the global data (acceptable within the 90\% C.L.)
with $\langle{x}\rangle_{\rm IC} \leq 1.5 \%$
for a sea-like IC in the SEA model at $Q_{c}=1.3\ {\rm GeV}$;
or with $\langle{x}\rangle_{\rm IC} \leq 2.5 \%$
for a valence-like IC (i.e., concentrated at large $x$) in the BHPS model, 
cf. Fig.~ \ref{fig:chisq+T2Vxic}.

The HERA combined data on inclusive charm production
in ep deep-inelastic scattering is particularly important.
It constrains the SEA model more strongly than other data.
On the other hand, it neither constrains nor favors
the large-$x$ IC of the BHPS model.

We also investigated the correlation between IC content
and the value of the charm quark mass in the global fits.
We found that a larger value of $m_c$ was better fit
with larger amounts of intrinsic charm, even for the SEA model.
(The best fit for $m_c=1.67$
GeV required about $\langle{x}\rangle_{\rm IC} \approx 1 \%$ for the SEA model.)
However, the HERA combined data on inclusive charm production
has a significant tension with larger values of the $m_c$,
which is only somewhat alleviated by including nonzero IC.
Thus, the overall best fit to the data with the SEA model
still appears be with the smaller charm quark mass $m_c=1.3$ GeV
and small or zero IC.

Predictions for LHC measurements, based on the IC PDFs, are interesting.
The figures in Sec.\ \ref{sec:PREDICTIONS} show that the IC models with
largest allowed IC, are right up
against the uncertainty limits for the CT10 PDFs.
In other words, the PDF \emph{uncertainties} are just as large as the
IC \emph{effects}.
Global analysis cannot say more about IC until the PDFs are more
accurately determined.
(Special interactions, especially sensitive to the charm quark but not
used in global analysis, might be able to say more.)
But more accurate measurements at the LHC will reduce the PDF uncertainties.
The results of Sec.\ \ref{sec:PREDICTIONS} show that new LHC results
will impact the limit on intrinsic charm.
Or, if $2\%$ intrinsic charm is \emph{real} then the results of
Sec.\ \ref{sec:PREDICTIONS} indicate that it will show up as
a discrepancy between theory and data in some common LHC measurements.
For example, Fig.\ 14
shows that the BHPS models can strongly modify the $p_{T,Z}$
distribution in associated production of $Z$ boson and charm jet.

Some recent research regarding intrinsic charm has
focused on inclusive charm production at the LHC or 
Tevatron~\cite{lhccharm,Kniehl09,Bugaev10,Lykasov12},
or on associated production of charm with a prompt 
photon~\cite{Abazov09,Stavreva09,Stirling12,Lipatov12,Bednyakov13}.
The IC PDFs constructed here can be used to make
up-to-date predictions for those processes.
This set of CT10 IC PDFs will be made available via an internet website.

\clearpage

\section*{Appendix: Spartyness: an equivalent gaussian variable
\label{sec:APPENDIX}}

We introduce the variable $S_{n}$,
to simplify the comparison of data and theory
for the many data sets in the global analysis.
$S_{n}$ is the equivalent \emph{Gaussian} variable,
equivalent in likelihood to a given \emph{chi-square} measure.\footnote{%
For a short nickname, we call $S_{n}$ the ``spartyness''.}

We could, of course, just use the individual values of
$\chi^{2}/N_{pt}$ as measures of the goodness-of-fit
for the different data sets,
e.g., as in Table \ref{tab:EXP_bin_ID}.
However, that variable has different meanings
for different values of $N_{pt}$.
For example, $\chi^{2}/N_{pt} = 11.0/10$ has a much different meaning
than $\chi^{2}/N_{pt} = 110.0/100$, although they have the same value.
The chi-square probability for $\chi^{2}/N_{pt} \geq 11.0/10$ is 0.358;
the probability for $\chi^{2}/N_{pt} \geq 110.0/100$ is 0.232.
And the probability for $\chi^{2}/N \geq 1100.0/1000$
is only 0.015.
Different data sets in the global analysis have very different
numbers of data $N_{pt}$;
so the meaning of different values of $\chi^{2}/N_{pt}$
is not manifest.

The variable $S_{n}$ is designed to clarify the goodness of fit,
by transforming the cumulative probability from
the Chi-Square distribution
(which can be misleading because it depends on $N$)
to the normal distribution (which is more familiar).

The spartyness $S_{n}$ is a function of $\chi^{2}$ and $N$.
Let $P(\chi^{2},N)$ denote the $\chi^{2}$-probability
distribution function (PDF) for $N$ variables.
Its cumulative distribution function (CDF) is
\begin{equation}
C(\chi^{2},N) = \int_{0}^{\chi^{2}} P(\xi,N) d\xi.
\end{equation}
The definition of $S_{n}$ is
\begin{equation}\label{eq:Spdef}
C(\chi^{2},N) =
\int_{-\infty}^{S_{n}}\ \frac{e^{-x^{2}/2} dx}{\sqrt{2\pi}}.
\end{equation}

We find that this variable is very helpful in judging the goodness
of fit for the individual experiments in the global analysis;
cf.\ Section 3.
The pure definition is not very convenient for computation.
Therefore we use an accurate approximation for $S_{n}$~\cite{TLewis}
\begin{eqnarray}
S_{n} & \approx &  L(\chi^{2},N_{pt}) \\
L &=& \frac{(18 N_{pt})^{3/2}}{18 N_{pt}+1}
\left\{\frac{6}{6-\ln(\chi^{2}/N_{pt})}-\frac{9 N_{pt}}{9N_{pt}-1}\right\}
\end{eqnarray}

\emph{Ideally,} the variable $S_{n}$ has an approximately Gaussian
distribution with mean 0 and standard deviation 1.
For a good fit to data, the value of $S_{n}$ should be roughly
between -1 and 1.
A fit with $S_{n} > 3$ should be considered a poor fit.
For a fit with $S_{n} < -3$, equivalent to $\chi^{2} \ll N_{pt}$,
we might assume that the actual systematic errors are smaller
than the values used in the calculation of $\chi^{2}$.

\emph{In reality,} the variable $S_{n}$ is unexpectedly large
for some data sets, which are never fit very well in global analysis.
Therefore, we do not judge the absolute values of $S_{n}$,
but rather the relative values---relative to the best global fit---when
comparing different sets of PDFs.
For example, in Figs.\ \ref{fig:spartyall} and \ref{fig:spartysig}
we assume it is not the absolute value of $S_{n}$ for a chosen experiment
that is relevant;
but the change in the value as a function of $\langle{x}\rangle_{\rm IC}$,
which informs about the sensitivity of the experiment to the
amount of intrinsic charm.

\paragraph{The Tier-2 penalty.}
When comparing the quality of agreement between theory and data,
especially comparing alternative PDFs to a central fit,
we impose additional \emph{penalties} for large increases of $S_{n}$.
The penalty $T_{2}(i)$ for experiment $i$ increases faster than
$\chi_{i}^{2}$ as distance to the central fit increases,
when the likelihood of data set $i$ goes below 10\%.
Then using $\chi^{2}+T_{2}$ to judge the goodness of fit,
we can exclude test PDFs for which one or a few data sets
do not respect the 90\% confidence level.
Technically, we define the Tier-2 penalty for experiment $i$
by $\left( S_{n}(i)/S_{n,{\rm best}}(i) \right)^{p}$,
where the exponent $p$ is greater than 2. The Tier 2 penalties
 in this  paper were calculated by arbitrarily taking $p$ to be 16.
(We do not impose an added penalty on
experiments with $S_{n}<0$.)

The Tier-2 penalty is not used in finding the central fit,
but it is used in constructing the error PDFs
in the Hessian error analysis. That helps to avoid error PDFs
that conflict (outside the 90\% confident level)
with individual data sets.
We also use the Tier-2 penalty when estimating uncertainties
of alternative PDFs in the manner of Figure 2.
The simple requirement that the global $\Delta\chi^{2}$
must be less than some chosen ``tolerance'' is not adequate;
we must assure that no individual experiment would definitively
rule out the alternative PDFs.

\clearpage

\section*{Acknowledgments}
 We thank G. Salam for the useful discussions on the Hoppet program, 
 and  M. Guzzi  and P. Nadolsky for discussions on the effect of 
 charm mass to the global analysis. 
 T.J. Hou thanks the hospitality of 
 the Michigan State University where part of his work was done. 
 This work was supported in part by the U.S. National 
 Science Foundation under Grant No. PHY-0855561; by the 
 U.S. DOE Early Career Research Award DE-SC0003870; and by the 
 National Natural Science Foundation of China 
 under the Grant No. 11165014.

\end{document}